\documentclass[preprint,groupedaddress,showpacs,superscriptaddress,amssymb,amsmath]{revtex4-1}
\usepackage{graphicx,epsfig,epstopdf}
\usepackage{color}
\usepackage{placeins}
\usepackage{hyperref}
\newcommand{\be}{\begin{equation}}
\newcommand{\ee}{\end{equation}}
\newcommand{\bea}{\begin{eqnarray}}
\newcommand{\eea}{\end{eqnarray}}
\color{black}
\begin{document}
\title{Evaluation of optical probe signals from  nonequilibrium systems}
\author{Bijay Kumar Agarwalla}
\affiliation{Department of Chemistry, University of California, Irvine, California 92617, USA}
\author{Konstantin E. Dorfman}
\affiliation{Department of Chemistry, University of California, Irvine, California 92617, USA}
\author{Shaul Mukamel}
\affiliation{Department of Chemistry, University of California, Irvine, California 92617, USA}
\date{\today}
\color{black}
\begin{abstract} 
We predict several effects associated with the optical response of systems prepared in a nonequilibrium state by impulsive optical excitations. The linear response depends on the phase of the electric field even if the initial nonequilirbium state has only populations, no coherences. Initial quantum coherences induce  additional phase dependence which also shows new resonances in nonlinear wave mixing. In systems strongly driven by an external optical field, the field frequency generates a phase dependent probe absorption.  This gives further control to manipulate the relative contribution to the linear signal due to initial populations and coherences.
\end{abstract}


\maketitle

\section{Introduction}

Spectroscopic signals are commonly expressed in terms of the polarization induced in the molecule by external driving \cite{Shaul5}. For $n$-wave mixing processes \cite{FWM_Mukamel, Shaul2}, the polarization is calculated perturbatively to $n\!-\!1$-th order in the field-matter interaction and written as  
a combination of $n$ point matter correlation functions which depend on the initial state of the molecule. In the frequency domain, the polarization is expressed in terms of the $n$-th order optical susceptibility $\chi^{(n)}$. For systems initially at equilibrium,  due to time-translational invariance symmetry, 
$\chi^{(n)}$ only depends on $n$ independent frequency variables. This is because the sum of all $n+1$ field frequencies must add up to zero. This reflects the energy conservation of the field. 

If the system can be prepared in an arbitrary nonequilibrium state the time-translational invariance symmetry is broken and the nonlinear optical response is governed by a more general $\chi^{(n)}$ that depends on $n+1$ independent frequency variables. This one extra frequency variable yields new resonances that are not present in the $n$-th order equilibrium response. To make a consistent connection with $n$-wave mixing results obtained with equilibrium state preparation, we introduce a generalized $n$-th order optical susceptibility $\tilde{\chi}^{(n)}$, which depends on the nonequilibrium initial state.
This initial state can contain populations and coherences.
Such nonequilibrium state prepared by an ultrashort actinic pulse is commonly studied in femtosecond Raman spectroscopy \cite{Miz97,Lee:JCP:2004,Kukura:AnnurevPhysChem:2007,Takeuchi:Science:2008}. In a different context, systems prepared in the nonequilibrium superposition of quantum states induced by strong monochromatic laser field have been  extensively studied in quantum optics \cite{Scully, Boyd} and quantum thermodynamics \cite{Boukobza, Geva, Scovil}. Monitoring field induced quantum coherence by the transmission of a weak probe shows many interesting effects including electromagnetic induced transparency \cite{Harris}, lasing without population inversion \cite{Lasing1,Lasing2} and coherent population trapping \cite{CPT1,CPT2}. Molecules in open junctions is another example of nonequilibrium preparation, where the molecule subjected to a chemical or thermal bias, across its boundaries, can be in a stationary state \cite{Boukobza}. For such initial conditions the standard fluctuation-dissipation relations are violated \cite{FD1,FD2,FD3}. In all the above examples nonequilibrium preparation affects the response of the system with respect to an optical perturbation. Typically the optical response of a system is probed by nonlinear $n$-wave mixing. For instance Raman $\tilde{\chi}^{(3)}$ measurements \cite{Kukura:AnnurevPhysChem:2007} can detect phase and amplitude information about resonances. In this paper we show that, similar information about phase can be obtained from simpler linear measurements described by $\tilde{\chi}^{(1)}$
which depends on two independent frequencies. The linear signal now depends on the phase of the electric field even if the system is initially prepared in a population state. The initial coherence further results in additional phase dependence which can be used in coherent control schemes \cite{control1, control2}. Another consequence of the nonequilibrium preparation are the new resonances in $n$ wave mixing. These are at the frequencies shifted from the usual $n$-wave mixing phase matching condition by an amount determined by the phase of the non stationary state corresponding to initial coherence \cite{Shaul1}. We further study an example where the initial conditions are considered explicitly for a three-level model systems driven by a strong monochromatic light source. We show how the control parameters of the driving field affect the phase dependence of the linear signal by manipulating initial coherences and populations.

The paper is organized as follows: In section II, we calculate linear absorption of systems initially prepared in an arbitrary superposition state and discuss the phase dependence due to the field envelope. Compact expressions are derived by employing superoperator notation.  In section III, we extend the calculation to nonlinear wave mixing signals and examine new resonances that show up because of the initial quantum coherence. In Section IV, we repeat the calculation for linear absorption of a strongly driven three level model system. The phase dependence now appears in the signal through the field frequency for both initial population and coherences states. We further examine the role of driving field parameters to control the relative contribution to the signal coming from population and coherences . In Section V, we summarize our findings and discuss possible experimental conditions where these effects can be realized.
\section{Phase dependence of linear optical signal}
We consider a multilevel quantum system driven by a classical
optical field and is described by the Hamiltonian  
\be
H(t) = H_{0}+H_{\rm int}(t),
\ee
where $H_{0}$ is the system Hamiltonian and $H_{\rm int}(t)$ is the light-matter interaction
\be
H_{\rm int} (t)= V \tilde{\cal E}(t)= V ({\cal E}(t)+{\cal E}^{*}(t)).
\ee
Here $V=\mu+\mu^{\dagger}$ is the dipole operator 
which is partitioned into lowering ($\mu$) and raising ($\mu^{\dagger}$)
operators and is responsible for the de-excitation and excitation between the molecular states,
respectively.
${\cal E}(t)$ and ${\cal E}^*(t)$ are positive and negative frequency components of the
total electric field $\tilde{\cal E}(t)={\cal E}(t)+{\cal E}^{*}(t)$. We will calculate
signals for an arbitrary initial molecular state. We
employ superoperator notation where with each Hilbert space operator
$A$, we associate two superoperators, denoted as $A_L$ (left) and $A_R$ (right), defined
through their action on Hilbert space operator $X$ as $A_L X \equiv A X,\, A_R X
\equiv X A$. We further define the linear combinations $A_{+}=(A_L
+ A_R)/2$
and $A_{-}= A_L - A_R$. A $+ (-)$ operation in Liouville space corresponds to an
anticommutation (commutation) operation in Hilbert space. Using this notation the frequency-dispersed
heterodyne detected signal is given by \cite{Rahav_stim, Rahav2_stim}
\be
S(\omega,\Gamma)= \frac{2}{\hbar}\,{\rm Im} \Big[ {\cal E}^*(\omega) \int_{-\infty}^{\infty}\, dt\, e^{i
\omega
(t-t_0)} \Big \langle {\cal T} V_L(t) \exp \big(-\frac{i}{\hbar} \int_{\tau_0}^{t} d\tau_1
H_{{\rm int}-}(\tau_1)  \big) \Big\rangle_{\rho(\tau_0)} \Big],
\label{first-definition}
\ee
where $\Gamma$ represents the set of parameters of the incoming fields. These will be specified later. The angular bracket $\langle \cdots \rangle$ represents an average over the initial density matrix of the molecule prepared at time $\tau_0$. The external fields have
finite envelopes centered at 
$t_0 \geq \tau_0$.  The superoperators in Eq.~(\ref{first-definition}) are in the interaction
picture with respect to the free system Hamiltonian $H_0$ i.e., for  any operator $A$,
$A_{\nu}(t)\equiv
\exp({i {H_0}_{-}(t-\tau_0)}) A_{\nu} \exp({-i {H_0}_{-}(t-\tau_0)}),$ $\nu=L,R$.
We further define the retarded Liouville space evolution operator
${\cal G}(t\!-\!\tau_0)= (-i/\hbar) \theta(t-\tau_0) \exp\big[\!-\!\frac{i}{\hbar} H_{0 -}(t-\tau_0)\big]$ and the advanced evolution operator ${\cal G}^{\dagger}(t\!-\!\tau_0)= (i/\hbar) \theta(\tau_0-t)
\exp\big[\!-\!\frac{i}{\hbar} H_{0 -}(t-\tau_0)\big]$. In the frequency domain these
propagators are written as
${\cal G}(\omega)=  \int_{-\infty}^{\infty} dt \,e^{i \omega (t-\tau_0)} {\cal G}(t-\tau_0)=
\frac{1}{\hbar}(\omega I -
\frac{1}{\hbar} H_{{0}-}+ i \epsilon)^{-1}$ and 
${\cal G}^{\dagger}(\omega)= \int_{-\infty}^{\infty} dt\, e^{-i \omega (t-\tau_0)} {\cal
G}^{\dagger}(t-\tau_0)=\frac{1}{\hbar}(\omega I -\frac{1}{\hbar}
H_{{0}-}-i\epsilon)^{-1}$. Here $I$ is the Identity operator in Liouville space.
${\cal T}$ is the time-ordering superoperator
\be 
{\cal T} A_{\nu} (t_1) B_{\nu'} (t_2)= \theta(t_1-t_2) A_{\nu}(t_1) B_{\nu'}(t_2) +
\theta(t_2-t_1) B_{\nu'}(t_2) A_{\nu}(t_1),  \quad \nu,\nu'=L,R.
\ee
The linear signal is obtained by expanding the
exponential in Eq.~(\ref{first-definition}) to
first order in the field-matter interaction $H_{\rm int -}$,
\be
S^{(1)}(\omega;t_0,\tau_0)= \frac{2}{\hbar}\,{\rm Im} \Big[i \hbar 
\int_{-\infty}^{\infty} dt
\int_{-\infty}^{t} d\tau_1 e^{i \omega (t-t_0)}\, {\cal E}^{*}(\omega)
\tilde{\cal E}(\tau_1) \,\,\big\langle V_L {\cal G}(t-\tau_1)
V_{-}{\cal G}(\tau_1-\tau_0)\big\rangle_{\rho(\tau_0)}
\Big].
\label{expand-first}
\ee
Note that ${\cal G}(\tau_1\!-\!\tau_0)= -\frac{i}{\hbar}{I}$ if $\rho(\tau_0)$ is a stationary
distribution (equilibrium or steady state). Otherwise it describes the evolution of the nonstationary initial state. The presence of
$\theta(\tau_1-\tau_0)$ in ${\cal G}(\tau_1-\tau_0)$ allows to extend the lower limit of integration 
for $\tau_1$ to $-\infty$. Making the change of
time variable $t-\tau_1= t_1$, we write 
\be
S^{(1)}(\omega;t_0,\tau_0)=  \frac{2}{\hbar}\,{\rm Im} \Big[i\hbar \int_{-\infty}^{\infty} dt
\int_{0}^{\infty} dt_1
e^{i \omega (t-t_0)} {\cal E}^{*}(\omega) \tilde{\cal E}(t\!-\!t_1\!-\!t_0)
\big\langle V_L {\cal G}(t_1) V_{-} 
{\cal G}(t-t_1-\tau_0)\big\rangle_{\rho(\tau_0)}
\Big]. 
\ee
Performing the inverse Fourier transformation for ${\cal
G}(t\!-\!t_1\!-\!\tau_0)$, $\tilde{\cal E}(t\!-\!t_1\!-\!t_0)$ and integrating over $t_1$ and $t$, we obtain 
\bea
S^{(1)}(\omega;t_0,\tau_0)&=&  \frac{2}{\hbar}\,{\rm Im} \Big[i\hbar \int_{-\infty}^{\infty}
\frac{d\omega_0}{2 \pi} \int_{-\infty}^{\infty} \frac{d\omega_1'}{2\pi}
e^{-i(\omega-\omega_1')t_0} e^{i \omega_0 \tau_0}\,{\cal E}^{*}(\omega) \tilde{\cal E}(\omega_1')
\big \langle V_L
{\cal
G}(\omega_1'+\omega_0) V_{-}{\cal
G}(\omega_0)\big \rangle_{\rho(\tau_0)}
\nonumber \\
&& \times\,
 2\pi \delta(\omega-\omega_0-\omega_1')\Big].
\eea
The $\omega_0$ integration can be carried out easily which results in  
\be 
S^{(1)}(\omega;t_0,\tau_0)=  \frac{2}{\hbar}\,{\rm Im} \Big[i\hbar \int_{-\infty}^{\infty}
\frac{d\omega_1'}{2\pi}\, e^{-i(\omega-\omega_1')(t_0-\tau_0)} \, {\cal E}^{*}(\omega)
\tilde{\cal E}(\omega_1') \big \langle V_L {\cal
G}(\omega) V_{-}{\cal G}(\omega-\omega_1')\big \rangle_{\rho(\tau_0)} \Big].
\label{final-linear-signal}
\ee
This expression holds for an arbitrary initial density matrix. 

We now assume that the molecule is prepared in a superposition of eigenstates $\rho(\tau_0)= \sum_{ab} \rho_{ab}|ab\rangle \rangle$ by a weak impulsive pulse centered at $\tau_0$ 
 and calculate the linear absorption of a second weak probe pulse. We neglect dephasing and assume that the probe interacts with the system while the coherence is alive. This is the case in recent experiments in photosynthetic complexes \cite{Pan10} and solar cells \cite{Pro14}. In Section IV we consider  a particular example where the strong driving field induces quantum coherence and it is long-lived.
For a superposition state $|ab\rangle\rangle$ the contour integration
over $\omega_1'$ in Eq.~(\ref{final-linear-signal}) can be carried out which gives
\bea
&&\int_{-\infty}^{\infty} \frac{d\omega_1'}{2\pi}  \tilde{\cal E}(\omega_1')  {\cal
G}_{ab}(\omega-\omega_1')e^{-i\omega_1'(\tau_0-t_0)} \nonumber \\
&& =-\frac{1}{\hbar}\lim_{\eta \to 0} \int_{-\infty}^{\infty} \frac{d\omega_1'}{2\pi}
\int_{-\infty}^{\infty} d\bar{t} \,\tilde{\cal E}(\bar{t})\,
\frac{e^{i\omega_1'(\bar{t}-\tau_0+t_0)}}{\omega_1'-(\omega-\omega_{ab})-i\eta}\nonumber
\\
&&=-\frac{i}{\hbar} \int_{-\infty}^{\infty} d\bar{t} \,  \,
\theta(\bar{t}-\tau_0+t_0)\,\tilde{\cal E}(\bar{t})\,
e^{i (\omega-\omega_{ab})(\bar{t}-\tau_0+t_0)},
\eea
where in the second line $\eta$ is an infinitesimal positive number used to satisfy the causality condition for the retarded propagator. In last line we use the definition of Heaviside theta function. The signal is finally given by  
\be 
S^{(1)}(\omega,;t_0-\tau_0)=  \frac{2}{\hbar}\,\sum_{ab}\,{\rm Im} \Big[\int_{-\infty}^{\infty} d\bar{t}
\, \theta(\bar{t}+t_0-\tau_0) \,{\cal E}^{*}(\omega) \tilde{\cal E}(\bar{t}) 
\big \langle V_L {\cal
G}(\omega) V_{-}\big \rangle_{\rho_{ab}}  e^{i (\omega-\omega_{ab}) \bar{t}} e^{-i \omega_{ab} (t_0-\tau_0)} \Big].
\label{final-linear-signal-1}
\ee
In the following we discuss various limits of the signal and its dependence on the phase of the field.

\subsection{Systems at equilibrium} 
The standard equilibrium result is recovered from Eq.~(\ref{final-linear-signal-1}) in the limit $\tau_0 \to -\infty$ where the molecule initially may reside in some population
state $|aa\rangle
\rangle$ (no coherence) or, in thermal equilibrium i.e., $\rho_{\rm eq}(\tau_0) =
e^{-\beta
H_0}/{Z}$, where  $\beta$ is the inverse temperature and $Z={\rm Tr}\big[e^{-\beta
H_0}\big]$ is the partition function. 
The linear signal then reduces to 
\be
S^{(1)}_{\rm eq}(\omega)= \frac{2}{\hbar} \,{\rm Im}\Big[{\cal E}^*(\omega) \tilde{\cal E}(\omega)
\chi^{(1)}(-\omega;\omega)\Big],
\label{eq-result}
\ee
where $\chi^{(1)}(-\omega;\omega)\equiv \langle V_L {\cal G}(\omega) V_{-} \rangle_{\rho_{eq}}$ which depends on a single frequency and
$\tilde{\cal E}(\omega)$ is the Fourier transformation of the total electric field $\tilde{\cal E}(t)$. Making the rotating wave approximation (RWA) and assuming that the molecule initially is in a ground
electronic state $|a\rangle$, we get for the linear signal 
\be
S^{(1)}_{\rm eq}(\omega)= \frac{2}{\hbar} \sum_{ac} {\rm Im} \Big[|{\cal E}(\omega)|^{2} \frac{|\mu_{ca}|^2 \rho_{aa}}{\omega-\omega_{ca}+i\eta}\Big].
\ee
This solely depends on the power spectrum of the field $|{\cal
E}(\omega)|^{2}$ and is independent of its phase.

\subsection{Nonequilibrium state and long pulses}
When the pulse envelopes are long (continuous wave (CW)) i.e., the field $\tilde{\cal E}(\bar{t})=\tilde{\cal E}\,e^{-i \omega_1 \bar{t}}$, and $\mathcal{E}(\omega)=2\pi\mathcal{E}\delta(\omega-\omega_1)$ the integrated signal (Eq.~(\ref{final-linear-signal-1})) is given as 
\be
S^{(1)} \equiv \int \frac{d\omega}{2\pi} S^{(1)}(\omega) = \frac{2}{\hbar} \sum_{ab} {\rm Im} \Big[ {\cal E}^{*} \tilde{\cal E} \langle V_L {\cal G}(\omega_1) V_{-} \rangle_{\rho_{ab}} {\cal G}_{ab}(\omega_1-\omega_1)\Big].
\ee
The signal is now independent of $t_0$ and $\tau_0$. Note that even if the system is initially prepared in a superposition state, i.e. there is nonzero initial coherence $\rho_{ab}$ the CW signal does not show $\omega_{ab}$ dependent resonances. 

\subsection{Initial nonequilibrium state with populations and coherenes}
Starting  with a general nonequilibrium state with initial populations and coherences and following Eq.~(\ref{final-linear-signal-1}) with the assumption that the field envelopes are centered at $t_0=\tau_0$ we obtain the linear signal 
\bea
S^{(1)}(\omega)&=& \frac{2}{\hbar}\,{\rm Im} \Big[\sum_{ab} {\cal E}^{*} (\omega) {\bar{\cal
E}}(\omega-\omega_{ab})\langle V_L {\cal G}(\omega) V_{-} 
\rangle_{\rho_{ab}} \Big], \nonumber \\
& \equiv& \frac{2}{\hbar}\,{\rm Im}  \Big[ \int \frac{d\omega_1'}{2\pi} {\cal E}^{*} (\omega) {\bar{\cal
E}}(\omega_1') \tilde{\chi}^{(1)}(-\omega,\omega_1')\Big],
\label{superposition}
\eea
where  
\be
\bar{\cal E}(\omega)= \int_{0}^{\infty} dt\, \tilde{\cal E}(t) \, e^{i \omega t},
\label{one-sided}
\ee
is the one-sided Fourier transform of the electric field which makes the signal dependent on the phase of the field. We define the generalized linear susceptibility 
\bea
\label{sus-linear}
\tilde{\chi}^{(1)}(-\omega,\omega_1')&=& \sum_{ab} \langle V_L {\cal G}(\omega) V_{-}
\rangle_{\rho_{ab}} \delta(\omega\!-\!\omega_1'\!-\!\omega_{ab}),
\eea
which depends on two independent frequency variables. Expansion of matter correlation function in molecular eigenstates gives
\be
\langle V_L {\cal G}(\omega) V_{-}
\rangle_{\rho_{ab}}= \sum_{c} \rho_{ab} \bar{\mu}_{ca} \Big[
\frac{\bar{\mu}_{bc}}{\omega-\omega_{cb} + i \eta}  -
\frac{\bar{\mu}_{cb}}{\omega-\omega_{ac}+ i\eta} \Big].
\label{sum-linear-coh}
\ee 
where ${\bar{\mu}}_{ac}=\mu_{ac}+\mu_{ac}^{*}$ is the
matrix element of the total dipole operator.

The generalized susceptibility in Eq.~(\ref{sus-linear}) may also be recast in terms of forward $(\cal{G})$ and backward $(\cal{G}^{\dagger})$ Liouville space propagators
\be
\tilde{\chi}^{(1)}(-\omega,\omega_1')= \frac{i}{2\pi} \sum_{ab} \rho_{ab} \langle \langle I| V_L {\cal G}(\omega) V_{-} \big[{\cal G}(\omega-\omega_1')-{\cal G}^{\dagger}(\omega-\omega_1')\big] |ab\rangle\rangle.
\label{chi1-GGdag}
\ee
In the $t_0=\tau_0$ limit we could also write the signal directly from Eq.~(8) as 
\be
S^{(1)}(\omega)=  \frac{2}{\hbar}\,{\rm Im} \Big[i\hbar \int_{-\infty}^{\infty}
\frac{d\omega_1'}{2\pi} \, {\cal E}^{*}(\omega)
\tilde{\cal E}(\omega_1') \big \langle V_L {\cal
G}(\omega) V_{-}{\cal G}(\omega-\omega_1')\big \rangle_{\rho(\tau_0)} \Big].
\label{alternate-linear-signal}
\ee
This equation contains an additional Green's function ${\cal G}(\omega-\omega_1')$ compared to systems initially at equilibrium. This frequency-domain Green's function is given by a one-sided Fourier transform of a corresponding time-domain Green's function in Eq. (6), which depends on a superoperator time ordering via the Heaviside theta-function. The latter is an intrinsic part of the bookkeeping of the field-matter interaction which is typically done by assigning the time-ordering operator to the matter correlation function, while allowing the fields to evolve in an unrestricted manner. The frequency domain electric field $\tilde{\mathcal{E}}(\omega_1')$ is given by a complete Fourier transform of the time-domain field envelope. Alternatively, the burden of time-ordering can be placed in the field correlation function while allowing matter to evolve in time without restriction. In this case the complete Fourier transform of the Green's function results in the delta-function as shown in Eq.(\ref{sus-linear}), while the electric field is transformed via one-sided Fourier transform as in Eq.~(\ref{superposition})-(\ref{one-sided}). Both representations are identical and the choice can be made according to the convenience in bookkeeping of the field-matter interactions and details of the signal. When time-translational symmetry is restored for an initial equilibrium or steady state (with or without coherence) the Green's function ${\cal G}(\omega-\omega_1')= \delta(\omega-\omega_1')$ and the linear signal or the susceptibility $\chi^{(1)}$ depend on to a  single frequency variable. The breakdown of time-translation symmetry due to initial nonequilibrium state gives rise to signal described by $\tilde{\chi}^{(1)}$ which depends on two frequency variables as shown in Eq.~(\ref{sus-linear}).

\subsubsection{Phase dependence induced by initial populations}
From Eq.~(\ref{superposition}) we separate the contribution to the signal from initial populations
\bea
S^{(1)}_{\rm pop}(\omega)&=& \frac{2}{\hbar}\,{\rm Im} \Big[\sum_{a} {\cal E}^{*} (\omega) {\bar{\cal
E}}(\omega)\langle V_L {\cal G}(\omega) V_{-}
\rangle_{\rho_{aa}} \Big].
\label{population}
\eea
Despite the fact that initial state contains only diagonal elements of the density matrix (populations), the signal in Eq.~(\ref{population}) depends explicitly on the phase of the electric field.
This is because the initial state at $\tau_0$ is not an equilibrium state which makes the signal depend on ${\cal {\bar E}}(\omega)$ instead of ${\cal E}(\omega)$. It is only in the limit $\tau_0 \to -\infty$ that the phase dependence disappears.

To demonstrate this phase dependence we consider a linearly chirped \cite{chirp, Debnath} Gaussian pulse with spectral phase $\phi(\omega)= \phi_0 + \phi''(\omega-\bar{\omega}_c)^2/2$. The electric field is given as 
\be
{\cal E}(\omega)= \sqrt{\pi} {\cal E}_0 T_0 e^{-(\omega-\bar{\omega}_c)^2 T_0^2/4} e^{i \phi'' (\omega-\bar{\omega}_c)^2/2},
\label{chirp-pulse}
\ee
where $T_0/\sqrt{2}$ is the field-transform-limited temporal width, $\delta\omega_L=\sqrt{2}/T_0$ is the spectral width, $\bar{\omega}_c$ is the central frequency of the field and $\phi''$ is the quadratic phase inducing the linear chirp. The corresponding temporal profile is
\be
{\cal E}(t) = \frac{{\cal E}_0}{2} \sqrt{\frac{\Gamma}{\Gamma_0}} e^{-\Gamma t^2} e^{-i \bar{\omega}_c t},
\ee
where $\frac{1}{\Gamma}= \frac{1}{\Gamma_0} - 2 i \phi''$ and $1/\Gamma_0=T_0^2$. The temporal width of the chirped pulse is given as $
T_p=T_0 \sqrt{1+ (2\phi''/T_0^2)^2}$ and the instantaneous frequency is $\omega(t)= \bar{\omega}_c + 2 \alpha t$
with $\alpha = (2 \phi'')/[T_0^4+ (2 \phi'')^2]$. Eq.~(\ref{one-sided}) is then given by 
\be
{\cal {\bar E}}(\omega)= \frac{\sqrt{\pi} {\cal E}_0 T_0}{4} e^{-(\omega-\bar{\omega}_c)^2 T_0^2/4} e^{i \phi'' (\omega-\bar{\omega}_c)^2/2} \Big(1+ i \,{\rm Erfi} \Big[\frac{\omega-\bar{\omega}_c}{2 \sqrt{\Gamma}}\Big]\Big),
\ee
where ${\rm Erfi}[z]$ is imaginary error function. It is then clear that the chirp rate dependence in the signal enters through ${\cal E}^{*}(\omega) {\cal {\bar E}}(\omega) \propto {\rm Erfi}[\frac{\omega-\bar{\omega}_c}{2 \sqrt{\Gamma}}]$ function. 

\subsubsection{Phase dependence due to initial coherence}
From Eq.~(\ref{superposition}) the contribution of initial coherences to the signal is 
\be
S^{(1)}_{\rm coh}(\omega)= \frac{2}{\hbar}\,{\rm Im} \Big[\sum_{a,b,a\neq b} {\cal E}^{*} (\omega) {\bar{\cal
E}}(\omega-\omega_{ab})\langle V_L {\cal G}(\omega) V_{-}
\rangle_{\rho_{ab}} \Big].
\label{coherence}
\ee
The signal now  shows a new resonance 
which depends on the initial coherence frequency
$\omega_{ab}$ between levels $a$ ans $b$. The field envelope depends on the shifted frequency ${\cal {\bar E}}(\omega-\omega_{ab})$ which also generate a phase dependent signal. For a chirped pulse, the chirp rate dependence in the signal is 
 ${\cal E}^{*} (\omega) {\bar{\cal
E}}(\omega-\omega_{ab}) \propto e^{i \phi''(\omega_{ab}^2 -2 (\omega-\bar{\omega}_c) \omega_{ab})}\,\,{\rm Erfi}[\frac{\omega-\bar{\omega}_c}{2 \sqrt{\Gamma}}]$.

\subsubsection{Application to a three level model system}
\begin{figure}
\includegraphics[width=0.95\columnwidth]{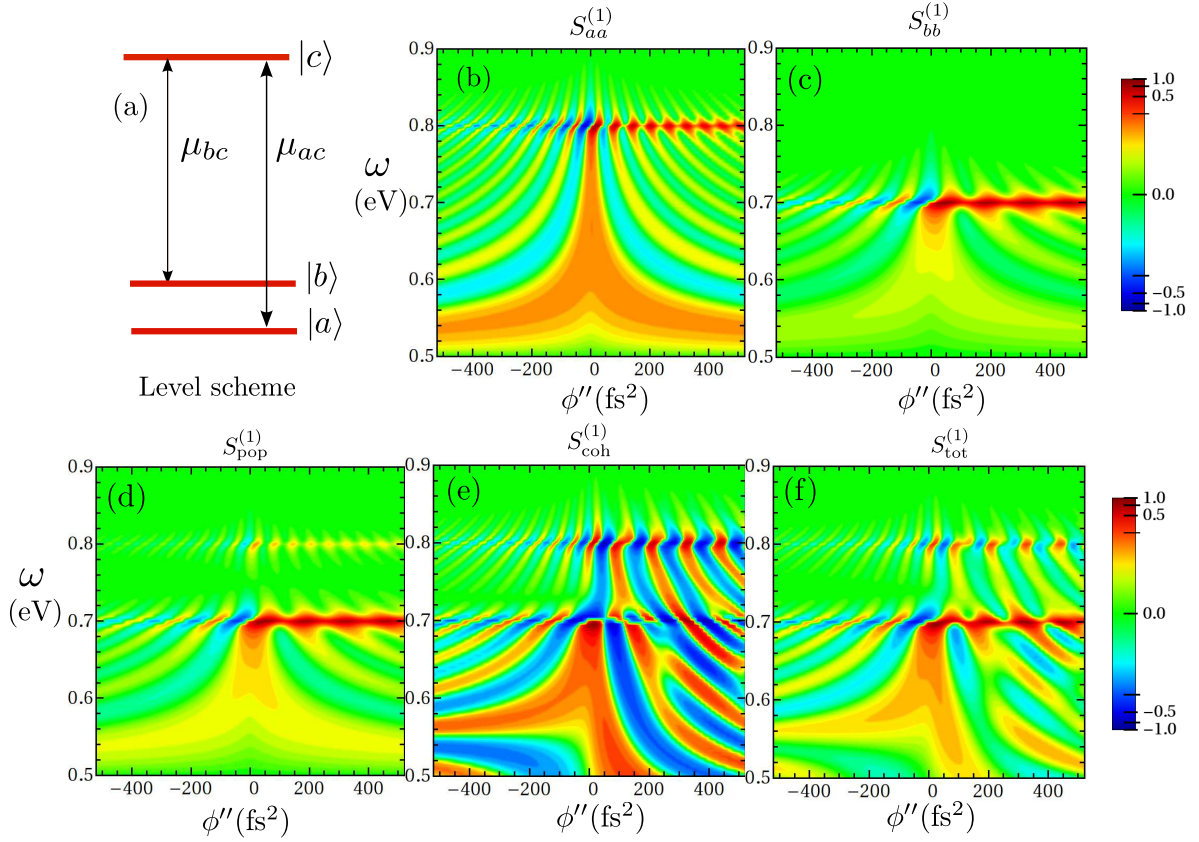}
\caption{(a) Level scheme for a three level model system with two lower states $|a\rangle$ and $|b\rangle$ and one upper state $|c\rangle$. The dipole transition is allowed between the states $|a\rangle \to |c\rangle$ and $|b\rangle \to |c\rangle$. The states $|a\rangle$ and $|b\rangle$ are initially in a maximally coherent state i.e., $\rho_{aa}\!=\!\rho_{bb}\!=\!\rho_{ab}\!=\!\rho_{ba}\!=\! \frac{1}{2}$. (b) 2D plot for the linear transmission signal $S_{aa}^{(1)}$ vs the detected frequency $\omega$ and chirp rate $\phi''$ starting with initial population in state $|a\rangle$, (c) Plot for $S_{bb}^{(1)}$ with initial population in state $|b\rangle$ and (d) the combined signal due to population $S_{\text{pop}}^{(1)}=S_{aa}^{(1)}+S_{bb}^{(1)}$. (e) The signal  $S_{\text{coh}}^{(1)}=S^{(1)}_{ab}+S^{(1)}_{ba}$ due to initial coherence ($\rho_{ab},\rho_{ba}$). (f) The total signal $S_{\text{tot}}^{(1)}$ including both population and coherences. The parameters are given as follows : $\omega_{ba}=0.1$ eV, $\omega_{ca}=0.8$ eV, $\bar{\omega}_c =0.5$ eV, $\eta=0.004$ eV, $T_0=6.6$ fs.}
\label{level-scheme-plots}
\end{figure}

In the following we demonstrate the phase dependence in the linear signal for the three level model system shown in Fig.~\ref{level-scheme-plots}(a). We assume that the two lower states $|a\rangle$ and $|b\rangle$ are initially in a maximally coherent state i.e., $\rho_{aa}\!=\!\rho_{bb}\!=\!\rho_{ab}\!=\!\rho_{ba}\!=\! \frac{1}{2}$ and all other elements of the density matrix vanish. Dipole transitions are allowed between states $|a\rangle \to |c\rangle$ and $|b\rangle \to |c\rangle$. Using Eq.~(\ref{superposition}) and making the RWA, the signal can be written as $S_{\text{tot}}^{(1)}(\omega)=\sum_{i,j=a,b} S_{ij}^{(1)}(\omega)$ where
\be
S_{ij}^{(1)}(\omega)= \frac{2}{\hbar} \,{\rm Im} \Big[{\cal E}^{*}(\omega) \bar{\cal E}(\omega-\omega_{ij}) \frac{\mu_{ci} \mu^{*}_{cj} \rho_{ij}}{\omega-\omega_{cj}+i \eta} \Big].
\ee
In Fig.~\ref{level-scheme-plots}(b) and Fig.~\ref{level-scheme-plots}(c) we display linear transmission vs the chirp rate and frequency when the system starts in population states $\rho_{aa}$ and $\rho_{bb}$ respectively. $S^{(1)}_{aa}$ and $S^{(1)}_{bb}$  show peaks corresponding to $\omega_{ca}=0.8$\,eV and $\omega_{cb}=0.7$\,eV transitions respectively. The population oscillations in the signal come from the error function which oscillates faster with the chirp rate $\phi''$ for higher $\omega-\bar{\omega}_c$. We use the central frequency $\bar{\omega}_c=0.5$ eV which implies that for $\omega_{ca}$ transition the oscillation is higher compared to $\omega_{cb}$ transition. Also the signal is asymmetric with respect to the chirp rate $\phi''$. In fact, it is stronger for positive $\phi''$ as compared to the negative one and is due to the amplitude of the field envelope which is higher for positive chirp.
In Fig.~\ref{level-scheme-plots} (d) we display the total signal due to initial populations $S_{\text{pop}}^{(1)}=S^{(1)}_{aa}+S^{(1)}_{bb}$. In Fig.~\ref{level-scheme-plots} (e) we plot the signal $S_{\text{coh}}^{(1)}=S^{(1)}_{ab}+S^{(1)}_{ba}$ due to initial coherence.  This signal appears to be stronger compared to that due to populations even for negative chirp. In Fig.~\ref{level-scheme-plots}(f) we display the total signal $S_{\text{tot}}^{(1)}$ due to initial population and coherence. 

\section{New resonances in nonlinear wave-mixing }
The Above results for the linear signal can be easily extended for nonlinear signals. In the following  we present expressions for three wave-mixing (TWM) and four-wave mixing (FWM) signals from an initial superposition state. The electric field $\tilde {\cal E}$ is a sum of two (for TWM) or three (for FWM) monochromatic field modes $\tilde{\cal E}_i,i=1,2,3$. Because all the frequency modes overlap in time, we must include all possible permutations of these modes to calculate the full signal.  We first write the TWM signal in terms of the total field $\tilde {\cal E}$
\bea
S^{(2)}(\omega)= \frac{2}{\hbar}\,{\rm Im}\Big[\int_{-\infty}^{\infty}
\frac{d\omega_1'}{2\pi} 
\int_{-\infty}^{\infty} \frac{d\omega_2'}{2\pi} {\cal E}^*(\omega) \tilde{\cal E}(\omega_1') 
\bar{\cal E}(\omega_2') 
\tilde{\chi}^{(2)}(-\omega,\omega_1',\omega_2')\Big],
\label{TWM-eq}
\eea
where the second-order generalized susceptibility is 
\be
\tilde{\chi}^{(2)}(-\omega,\omega_1',\omega_2') \!=\! \sum_{ab} \langle V_L {\cal
G}(\omega) V_{-} {\cal G}(\omega-\omega_1') V_{-} 
\rangle_{\rho_{ab}} \, \delta(\omega\!-\!\omega_1'\!-\!\omega_2'\!-\!\omega_{ab}).
\ee
This can be alternatively recast as
\be
\tilde{\chi}^{(2)}(-\omega,\omega_1',\omega_2') \!=\! \frac{i}{2\pi} \sum_{ab} \rho_{ab} \langle \langle I| V_L {\cal G}(\omega) V_{-} {\cal G}(\omega-\omega_1') V_{-} \big[{\cal G}(\omega\!-\!\omega_1'\!-\!\omega_2')-{\cal G}^{\dagger}(\omega\!-\!\omega_1'\!-\!\omega_2')\big] |ab\rangle\rangle.
\ee
Expanding the matter correlation function in system eigenstates gives 
\bea
\langle V_L {\cal
G}(\omega) V_{-} {\cal G}(\omega-\omega_1') V_{-}  \rangle_{\rho_{ab}} &=& \sum_{cd}\! \rho_{ab}\!
\Bigg[\bar{\mu}_{dc} \bar{\mu}_{ca} {\cal G}_{cb}(\omega-\omega_1')
\bigg(\bar{\mu}_{bd} {\cal G}_{db}(\omega)-\bar{\mu}_{db} {\cal
G}_{cd}(\omega)\bigg) \nonumber \\
&+& \bar{\mu}_{da} \bar{\mu}_{cb} {\cal G}_{ac}(\omega-\omega_1')
\bigg(\bar{\mu}_{dc} {\cal G}_{ad}(\omega)-\bar{\mu}_{cd} {\cal
G}_{dc}(\omega)\bigg)\Bigg].
\eea
Extending these results to FWM we can similarly write 
\bea
S^{(3)}(\omega)= \frac{2}{\hbar}\,{\rm Im}\Big[\int_{-\infty}^{\infty}
\frac{d\omega_1'}{2\pi} 
\int_{-\infty}^{\infty} \frac{d\omega_2'}{2\pi} \int_{-\infty}^{\infty}
\frac{d\omega_3'}{2\pi} {\cal E}^*(\omega) \tilde{\cal E}(\omega_1') \tilde{\cal E}(\omega_2') \bar{\cal E}(\omega_3')
\tilde{\chi}^{(3)}(-\omega,\omega_1',\omega_2',\omega_3')\Big], \quad \quad
\label{FWM-eq}
\eea
where
\bea
\tilde{\chi}^{(3)}(-\omega,\omega_1'\omega_2',\omega_3')\!&=&\! \sum_{ab}\! \langle V_L {\cal
G}(\omega) V_{-} {\cal G}(\omega\!-\!\omega_1') V_{-} {\cal
G}(\omega\!-\!\omega_1'\!-\!\omega_2') V_{-} \rangle_{\rho_{ab}} \,\delta(\omega \!-\!\omega_1'-\!\omega_2'-\!\omega_3'-\!\omega_{ab}), \nonumber \\
&=& \frac{i}{2\pi} \!\sum_{ab} \rho_{ab} \langle \langle I| V_L {\cal G}(\omega) V_{-} {\cal G}(\omega\!-\!\omega_1') V_{-} {\cal G}(\omega\!-\!\omega_1'\!-\!\omega_2') V_{-} \nonumber \\
&& \big[{\cal G}(\omega\!-\!\omega_1'\!-\!\omega_2'\!-\!\omega_3')\!-\!{\cal G}^{\dagger}(\omega\!-\!\omega_1'\!-\!\omega_2'-\omega_3')\big] |ab\rangle\rangle.
\eea
Expanding in eigenstates finally gives 
\begin{equation}
\begin{split}
&\langle V_L {\cal
G}(\omega) V_{-} {\cal G}(\omega-\omega_1') V_{-} {\cal
G}(\omega\!-\!\omega_1'\!-\!\omega_2') V_{-} \rangle_{\rho_{ab}} = \nonumber \\
& \sum_{cde}\! \rho_{ab}\!
\Bigg[\bar{\mu}_{ca} \bar{\mu}_{dc} \bar{\mu}_{be} \bar{\mu}_{ed}\, {\cal
G}_{cb}(\omega\!-\!\omega_1'\!-\!\omega_2') \, {\cal G}_{db}(\omega\!-\!\omega_1')\,
\Big({\cal
G}_{eb}(\omega)- {\cal
G}_{de}(\omega)\Big) \\
&+
\bar{\mu}_{ca} \bar{\mu}_{bd} \bar{\mu}_{ec} \bar{\mu}_{de} \,{\cal
G}_{ad}(\omega\!-\!\omega_1'\!-\!\omega_2')\,
{\cal G}_{cd}(\omega\!-\!\omega_1') \, \Big({\cal
G}_{ed}(\omega)-{\cal G}_{ce}(\omega)\Big)\\
& +\bar{\mu}_{ec}
\bar{\mu}_{de} \bar{\mu}_{bd} \bar{\mu}_{ac} \, {\cal
G}_{cb}(\omega\!-\!\omega_1'\!-\!\omega_2')\, {\cal G}_{cd}(\omega\!-\!\omega_1') \Big({\cal
G}_{ed}(\omega)-{\cal G}_{ce}(\omega)\Big)\, \\
&\!+\! \bar{\mu}_{ea}
\bar{\mu}_{de} \bar{\mu}_{cd} \bar{\mu}_{bc} \,{\cal
G}_{ac}(\omega\!-\!\omega_1'\!-\!\omega_2')\,
{\cal G}_{ad}(\omega\!-\!\omega_1')\, \Big({\cal
G}_{ae}(\omega)- {\cal
G}_{ed}(\omega)\Big)\Bigg].
\end{split}
\label{sum-nonlinear-coh}
\end{equation}
The equilibrium limit for TWM and FWM signals can be obtained in a similar way as was done in Eq.~(\ref{eq-result}).

Below we present the FWM signal for the three-level model system, shown in Fig~(\ref{level-scheme-plots}a). The complex field amplitude  contains three monochromatic  modes and a spectrally broad Gaussian probe field. ${\cal E}(t)= \sum_{j=1}^3 {\cal E}_j \exp(i {\bf k}_j r_j -i \omega_j t) + \int \frac{d\omega}{2\pi} {\cal E}_4 (\omega) \exp(i {\bf k} r - i \omega t)$ where the probe field ${\cal E}_4(\omega)=\sqrt{\frac{2 \pi}{\sigma}} \exp\big[{(\omega-\bar{\omega_c})^2/2 \sigma^2}\big]$. We select the following phase matching direction for the signal ${\bf k}= {\bf k}_1\!-\!{\bf k}_2\!+\!{\bf k}_3$. The signal is calculated using Eq.~(\ref{FWM-eq}). As before we assume that the system is in a maximally coherent state. 
Various Liouville space pathways and corresponding expressions for the FWM signal are given in Appendix A. In Fig.~\ref{figure-FWM}(a) we display the signal starting with a population $S^{(3)}_{\rm pop}(\omega)$ vs the detuning  $\Delta=\omega\!-\!\omega_1\!+\!\omega_2\!-\!\omega_3$. Various peaks correspond to the different quantum pathways of matter represented by frequency-domain propagators, which depend on  some frequency combinations of the incoming field. The $\Delta=0$ peak corresponds to a detected frequency $\omega=1.35$ eV and is marked by an arrow. For the system initially prepared in a coherence the original peak at $\Delta=0$ splits into two peaks at $\Delta= \pm \omega_{ab}$, as marked by two arrows in Fig.~\ref{figure-FWM}(b). The spectra in (a) and (b) are not identical, as different matter pathways contribute to the signal if the system is initially prepared in a population or in coherence. In Fig.~\ref{figure-FWM}(c) we compare the total signal $S^{(3)}_{\rm tot}(\omega)=S^{(3)}_{\rm pop}(\omega) + S^{(3)}_{\rm coh}(\omega)$ with the signal coming only due to the population $S^{(3)}_{\rm pop}(\omega)$. The effect of initial coherence in the total signal is significant and many single photon resonances appear much stronger in the presence of coherence. In Fig.~\ref{figure-FWM}(d) we compare the signal obtained for a system initially prepared in a non equilibrium population state vs equilibrium state. Note that the data for the population is scaled.  The signal contain the same number of peaks but with different magnitude. The difference is solely due to the non equilibrium preparation of the initial state. The two spectra coincide in the limit $\tau_0 \to-\infty$ whereby $\bar{\mathcal{E}}\to\tilde{\mathcal{E}}$.
\begin{figure}
\includegraphics[width=0.95\columnwidth]{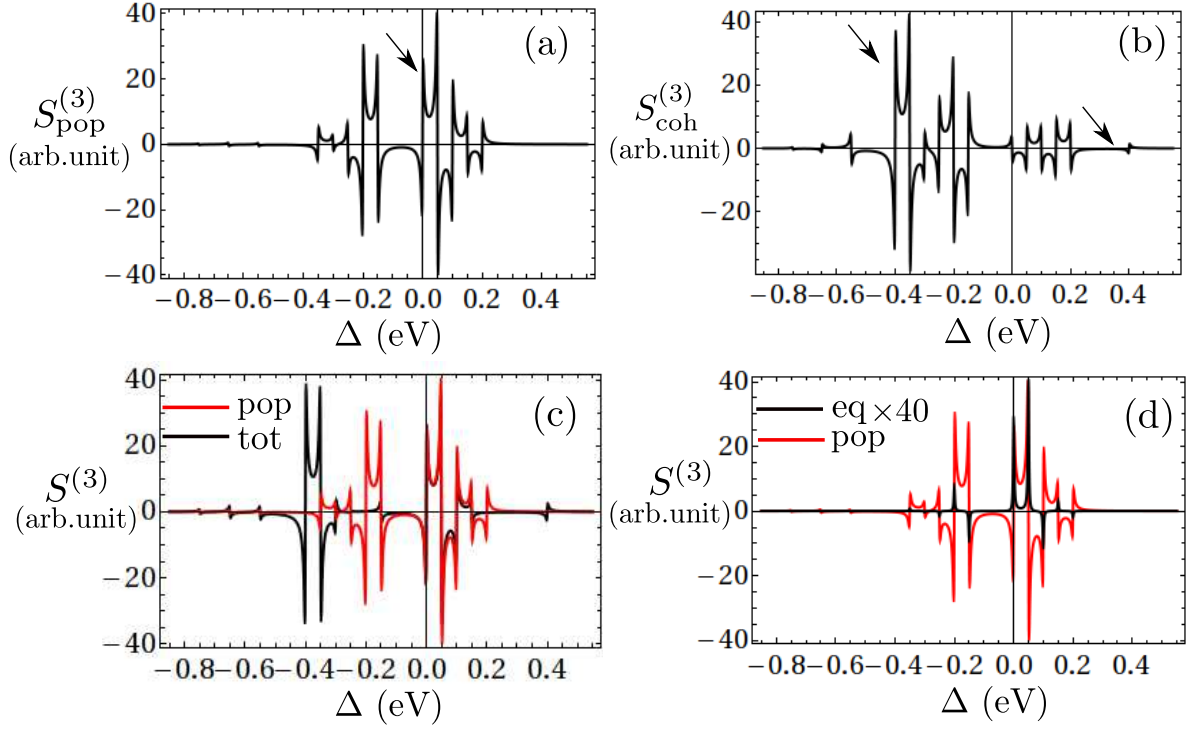}
\caption{(Color online) Plot for four FWM signal vs the detuning from the phase matching $\Delta=\omega-\!\omega_1+\!\omega_2-\!\omega_3$ for different initial conditions. $S^{(3)}_{\rm pop}$ - (a), $S^{(3)}_{\rm coh}$- (b), $S^{(3)}_{\rm total}$ (black) and  $S^{(3)}_{\rm pop}$(red) - (c), and $S^{(3)}_{\rm pop}$ (black) and $S^{(3)}_{\rm eq}$ (red) - (d). A single arrow in (a) and two arrows in (b) correspond to the splitting of a singe peak in population in two new resonant peaks in the presence of coherence as $\Delta=0$ is replaced by $\Delta=\pm \omega_{ab}$. The parameters are given as follows: $\omega_a=0$ eV, $\omega_b=0.4$ eV, $\omega_c=1.2$ eV, $\omega_1=1.1$ eV, $\omega_2=0.75$ eV, $\omega_3=1.0$ eV, $\bar{\omega}_c=0.5$ eV, $\sigma=10$ eV, $\eta=0.002$ eV.}
\label{figure-FWM}
\end{figure}
\FloatBarrier


\section{Linear response of a strongly driven system}
\subsection{Dressed state description for a strongly driven three-level system}
\begin{figure}
\includegraphics[width=0.3\columnwidth]{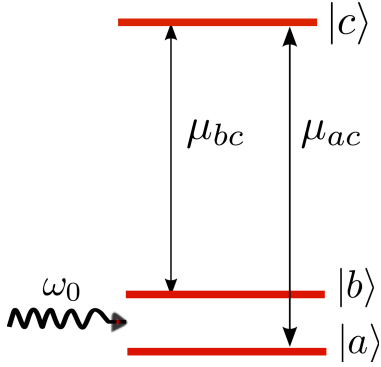}
\caption{Level scheme: Three level system with two lower states $|a\rangle$ and $|b\rangle$ and one upper state $|c\rangle$. The states $|a\rangle$ and $|b\rangle$ are driven by strong monochromatic field with frequency $\omega_0$. The optical transition is allowed between levels $|a\rangle \to |c\rangle$ and $|b\rangle \to |c\rangle$.}
\label{level-scheme}
\end{figure}

So far we did not specify how the system has been prepared in the stationary superposition of quantum states. We now consider a specific type of preparation using a strong driving field.  We examine the linear absorption of a three level system, shown in Fig.~(\ref{level-scheme}),
which is first prepared in a non-stationary state by driving the two lower levels  $|a\rangle$ and $|b\rangle$ with a strong monochromatic field of frequency $\omega_0$. 
The system is further in contact with a thermal bath that causes dephasing and relaxation. We consider a linear measurement taking this non-stationary state as an initial state.
The system plus bath is described by the Hamiltonian
\bea
H(t)&=&H_0(t)+H_B + H_{SB},\nonumber \\
H_0(t)&=&  \sum_{i=a,b,c} \hbar \omega_i |i\rangle \langle i|
-\frac{\mu {\cal E}_0}{2} \big[e^{i\omega_0 t} |a\rangle \langle
b| + e^{-i\omega_0 t}|b \rangle \langle a|\big], \nonumber\\
H_B &=& \sum_{k} \hbar \omega_k a_k^{\dagger} a_k, \nonumber \\
H_{SB} &=& \hbar \sum_{i\neq j=a,b,c,i<j}  a_k^{\dagger} |i\rangle \langle j| + h.c. ,
\label{Driven-Bloch-1}
\eea
where $H_0(t)$ is the Hamiltonian for the system including the interaction due to the driving field. $a_k (a_k^{\dagger})$ is the Bosonic annihilation (creation) operator for the bath, represented by the Hamiltonian $H_{B}$ and $H_{SB}$ is the system-bath coupling Hamiltonian. The driving electric field ${\cal E}_d(t)={\cal E}_0 \cos(\omega_0 t)$ and $\omega_0$
is near
resonant with the low energy states $|a\rangle$ and $|b\rangle$. The Schr\"odinger picture
evolution of the reduced density matrix in the lab frame 
is given by the Bloch equations
$\dot{\rho}(t)= {\cal L}(t) \rho(t)$ (see Appendix B). 
The driving field couples the
population of the two low energy states i.e., ${\rho}_{aa}$ and ${\rho}_{bb}$ with its coherence ${\rho}_{ab}$ and maintains a finite long-time coherence. 
\begin{figure}
\includegraphics[width=0.65\columnwidth]{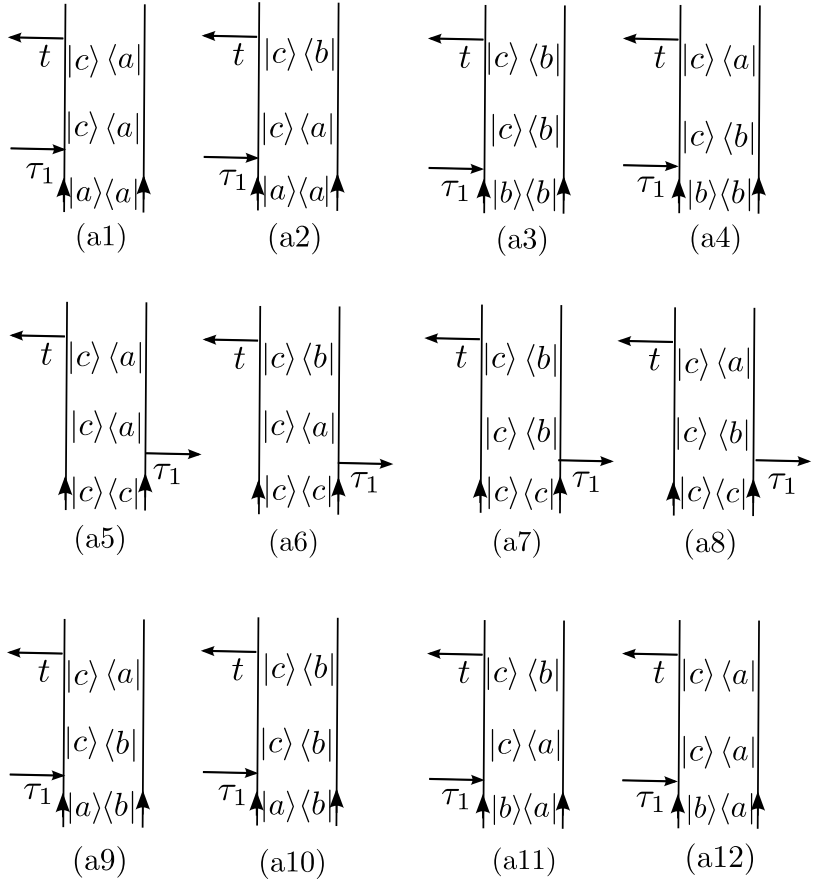}
\caption{Ladder diagrams for a three level model system driven by a monochromatic field. Diagrams
[(a1)-(a8)] are for initial conditions starting with population states. Diagrams [(a9)-(a12)]
are for initial conditions with coherence states. Time is increasing from bottom to top.}
\label{ladder-diagrams}
\end{figure}

We shall probe this system prepared in a nonstationary state at time $\tau_0$ with a weak optical field
${\cal E}(t)$ which allows transitions between the states $|a\rangle \to |c\rangle$ and $|b\rangle \to |c\rangle$. We write the light-matter interaction Hamiltonian in the RWA as
\be
H_{\rm int}(t)= {\cal E}(t) V^{\dagger}+{\cal E}^*(t) V
\ee
where $V= \sum_{i=a,b} \mu_{ic} |i\rangle \langle c|$ is now the de-excitation operator .

The contribution to the signal comes from various Liouville space pathways starting with populations and
coherences.  Corresponding expressions can
be obtained from the 
diagrams shown in the Fig.~(\ref{ladder-diagrams}) following standard diagrammatic rules \cite{Harbola}.  The total signal is $S^{(1)}_{\rm tot}(\omega;\omega_0,\Omega)=S^{(1)}_{\rm
pop}(\omega;\omega_0,\Omega)+S^{(1)}_{\rm coh}(\omega;\omega_0,\Omega)$. It depends parametrically on the optical frequency $\omega_0$ and the Rabi frequency $\Omega=\mu{\cal E}_0/2\hbar$ of the driving field. For  a system initially prepared in a population state (diagrams (a1)-(a8) in Fig.~4)) we obtain (see Appendix B) 
\bea
S^{(1)}_{\rm pop}(\omega;\omega_0,\Omega)&=&\frac{2}{\hbar}\,(\tilde{\rho}_{aa}^{ss}-\tilde{\rho}_{cc}^
{ ss } ) {\rm Im}
\Bigg[{\cal
 E}^*(\omega)
 \Bigg({\cal E}(\omega) \mu_{ac} \mu_{ca} \tilde{\cal G}_{ca;ca}(\omega\!-\!\omega_0) +
{\cal E}(\omega\!+\!\omega_0) \mu_{bc} \mu_{ca} \tilde{\cal
G}_{cb;ca}(\omega)\Bigg)\Bigg] \nonumber \\
&+& \frac{2}{\hbar}\,(\tilde{\rho}_{bb}^{ss}-\tilde{\rho}_{cc}^{ss}) {\rm Im} \Bigg[{\cal
E}^*(\omega)
\Bigg({\cal E}(\omega)\mu_{bc} \mu_{cb}
\tilde{\cal G}_{cb;cb}(\omega) +{\cal E}(\omega\!-\!\omega_0) \mu_{ac} \mu_{cb}
\tilde{\cal G}_{ca;cb}(\omega\!-\!\omega_0)\Bigg)\Bigg], \quad \quad 
\label{eq:population}
\eea
and for initial coherences (diagrams (a9)-(a12) in Fig.~4) we get
\bea
S^{(1)}_{\rm coh}(\omega;\omega_0,\Omega)&=& \frac{2}{\hbar}\,{\rm Im} \Bigg[{\cal
E}^*(\omega)
\Bigg(\big({\cal
E}(\omega\!+\!\omega_0)\mu_{bc} \mu_{ca} \tilde{\cal
G}_{cb;cb}(\omega)+ {\cal E}(\omega)\mu_{ac} \mu_{ca} \tilde{\cal
G}_{ca;cb}(\omega\!-\!\omega_0)\big)\tilde{\rho}_{ab}^{ss} \nonumber \\ 
&+&\big({\cal
E}(\omega-\omega_0) \mu_{ac} \mu_{cb}  
 \tilde{\cal G}_{ca;ca}(\omega\!-\!\omega_0)+ {\cal
E}(\omega) \mu_{bc} \mu_{cb} \tilde{\cal
G}_{cb;ca}(\omega)\big)\tilde{\rho}_{ba}^{ss} \Bigg)\Bigg],
\label{eq:coherence}
\eea
where the propagator $\tilde{\cal G}_{kl;mn}(\omega)= \langle \langle kl| (\omega I- i \tilde{\cal
L})^{-1}|
mn \rangle \rangle$ and $I$ is the identity matrix in Liouville space. $\tilde{\cal L}$ is defined in Appendix B. In the expression for the signal the $\omega_0$ dependence enters through the field 
as well as the retarded propagator. The presence of $\omega_0$ makes the signal phase dependent.
When $\omega_0=\Omega=0$ we recover the equilibrium result (diagrams
(a1),(a3), (a5) and (a7)) and the signal reduces to 
\be
S^{(1)}_{\rm eq}(\omega)=\frac{2}{\hbar}\, {\rm Im} \Big[|{\cal E}(\omega)|^{2} \sum_{i=a,b}
\big(\rho_{ii}^{ss}-\rho_{cc}^{ss}\big) \mu_{ic}\mu_{ci}
\tilde{\cal G}_{ci;ci}(\omega)\Big],
\ee
which solely depends on the power spectrum of the field and is independent of its phase. 

\subsection{Simulations of the linear response of a strongly driven three-level system}
\begin{figure}[t]
\includegraphics[trim=0cm 0cm 0cm 0cm, angle=0, width=0.95\columnwidth]{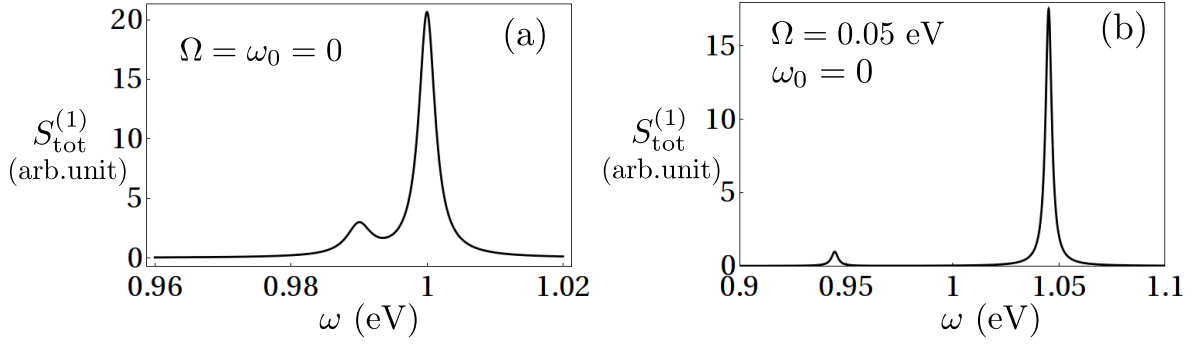}
\caption{Linear transmission signal $S^{(1)}(\omega)$ for a strongly driven three level model system as a function of detected frequency $\omega$ for two different cases. (a) - no driving $\Omega=0$, $\omega_0=0$, (b) - static coupling: $\omega_0=0$, $\Omega=0.05$ eV. Other parameters are: $\omega_a=0$ eV, $\omega_b=0.01$ eV, $\omega_c=1.0$
 eV, $k_{B}T= 0.0259$ eV, $\gamma_{ba}=0.004$ eV, $\gamma_{ca}=0.0001$ eV,
$\gamma_{cb}=0.0002 $ eV. Parameters for electric field are $T_0=0.14$ fs, $\bar{\omega}_c= 0.5$ eV, $\phi''=0$.}
\label{equilibrium}
\end{figure}

\begin{figure}[t]
\includegraphics[width=0.6\columnwidth]{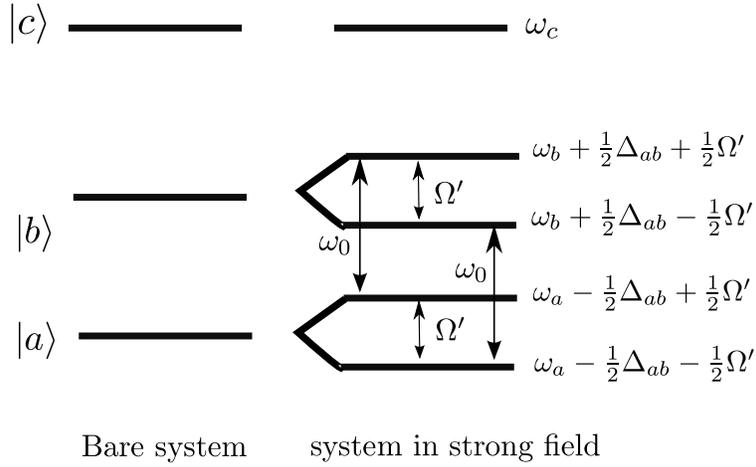}
\caption{(Color online) Frequency spectrum for the three level system with lower two levels driven by a strong field with frequency $\omega_0$. Here $\Delta_{ab}=\omega_0-\omega_{ba}$ and $\Omega'= (4 \Omega^2 + \Delta_{ab}^2)^{\frac{1}{2}}$. For details see Ref.~\cite{Boyd}}
\label{dressed}
\end{figure}

\begin{figure}[t]
\includegraphics[trim=0cm 0cm 0cm 0cm, angle=0, width=0.95\columnwidth]{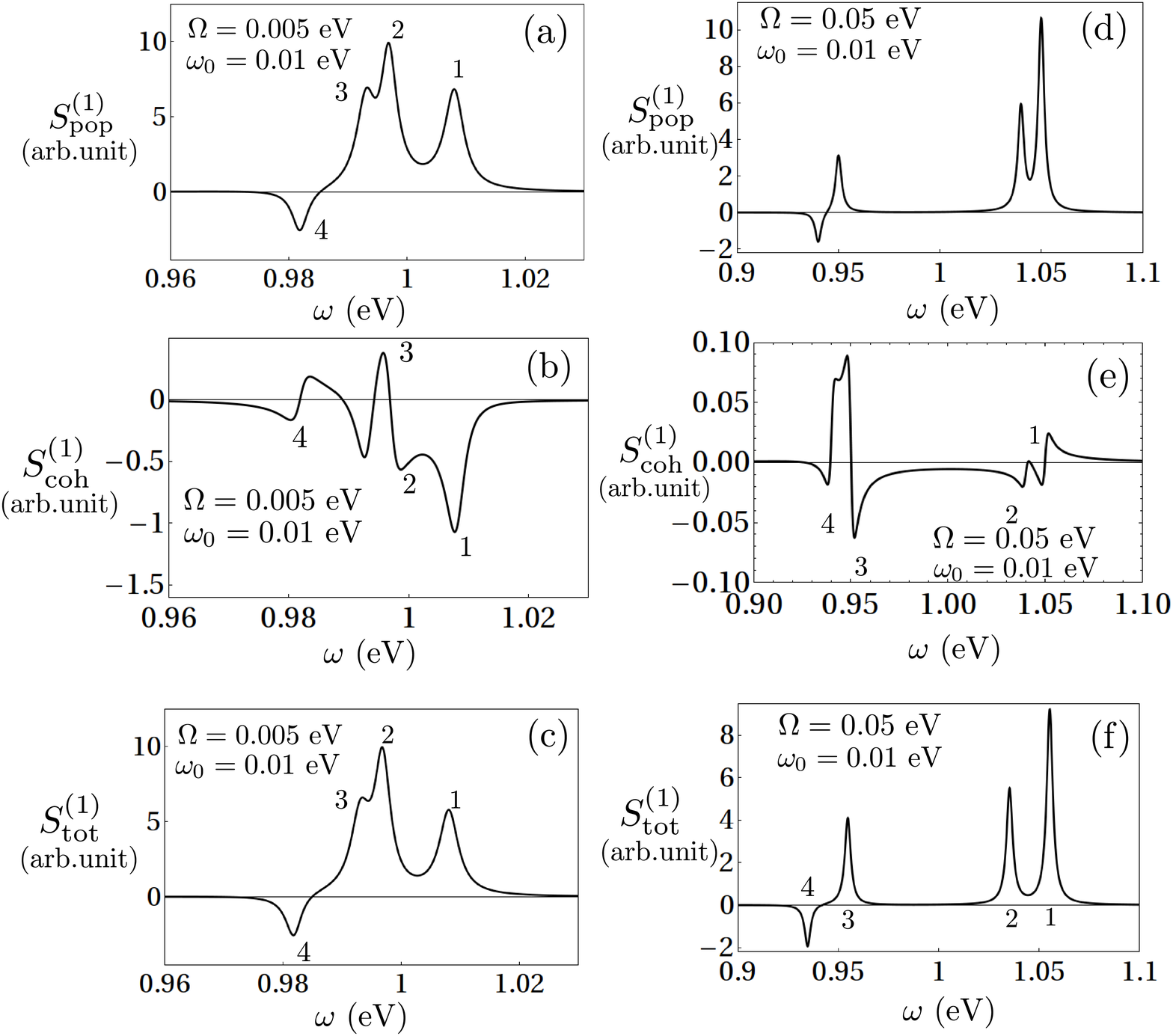}
\caption{Left column: $S_{\text{pop}}^{(1)}(\omega)$ - (a), $S_{\text{coh}}^{(1)}(\omega)$ - (b), $S_{\text{tot}}^{(1)}(\omega)$ - (c) for moderate driving $\Omega=0.005$ eV, $\omega_0=0.01$ eV. Right column: the same as the left but for the strong driving $\Omega=0.05$ eV, $\omega_0=0.01$ eV. Other parameters are the same as in Fig. \ref{equilibrium}}.
\label{bloch_signal2}
\end{figure}

\begin{figure}[t]
\includegraphics[width=0.9\columnwidth]{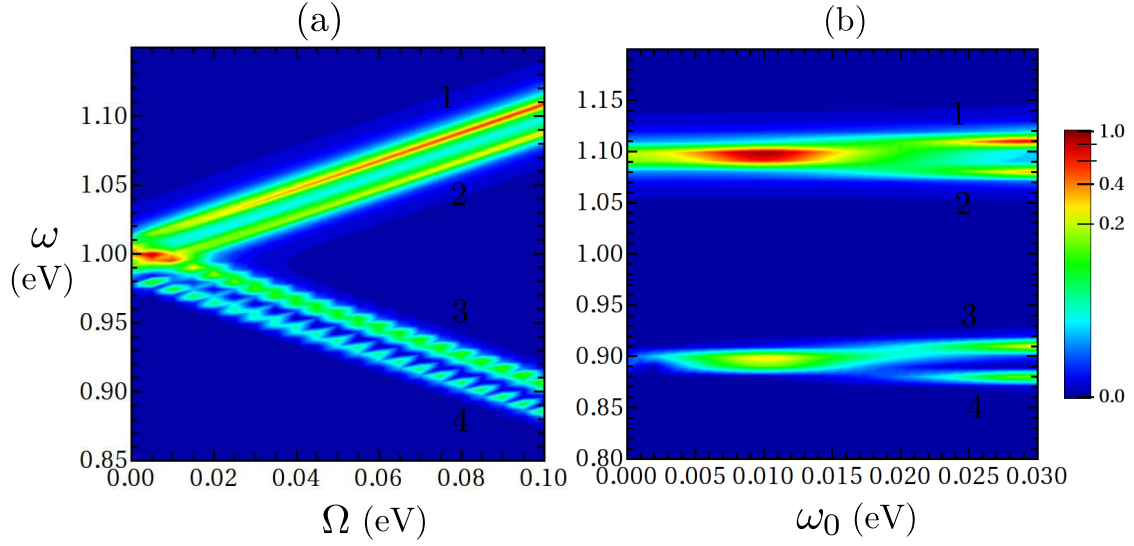}
\caption{(Color online) (a) absolute value of linear transmission signal $|S^{(1)}(\omega)|$ for three level
model system vs $\omega$ and $\Omega$ at fixed $\omega_0=0.02$ eV. (b) - the same signal but vs $\omega$ and $\omega_0$ at fixed $\Omega=0.1$ eV. Other parameters are the same as in Fig.~\ref{equilibrium}
}
\label{contour}
\end{figure}

\begin{figure}[t]
\includegraphics[width=1.05\columnwidth]{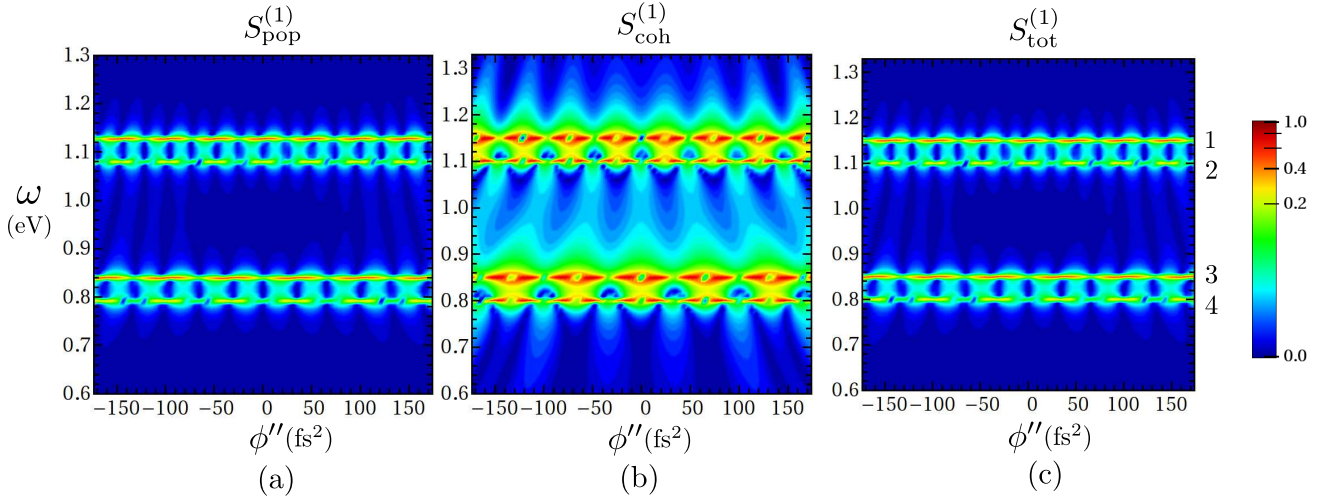}
\caption{(Color online) (a) 2D contour plot of absolute value linear transmission signal $|S^{(1)}(\omega)|$ vs detected frequency $\omega$ and chirp rate $\phi''$ starting with initial population, (b) starting with coherence and (c) the total signal. The values of the parameters are $\omega_{ba}=0.05$ eV, $\omega_{ca}=1.0$ eV, $\omega_0=0.05$ eV, $\Omega=0.15$ eV, $\bar{\omega}_c=0.3$ eV. Other parameters are the same as in  Fig.~\ref{equilibrium}.}
\label{contour-phase}
\end{figure}

\begin{figure}[t]
\includegraphics[width=1.05\columnwidth]{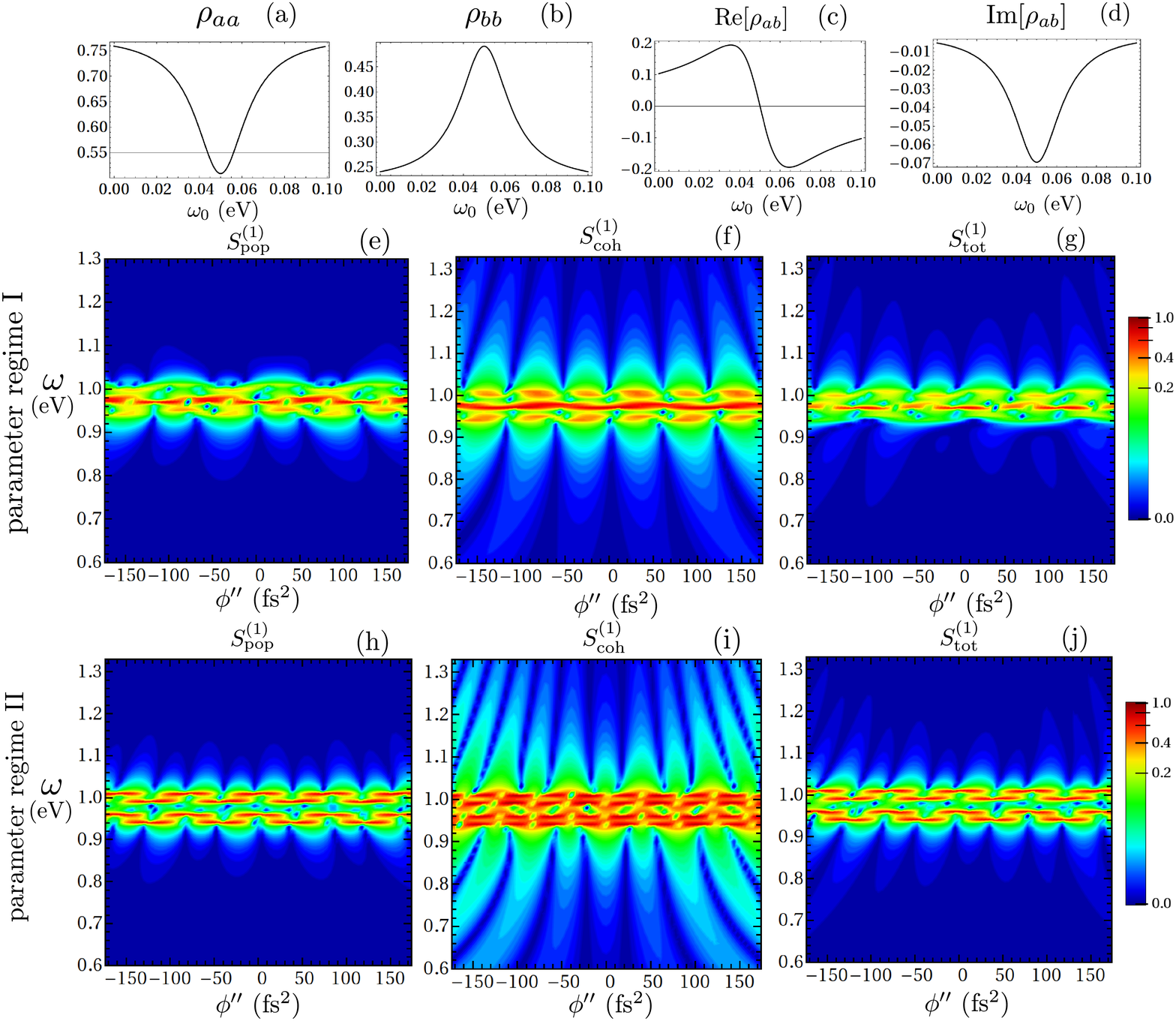}
\caption{(Color online) First row: density matrix elements vs $\omega_0$ for $\Omega=0.01$ eV: population $\rho_{aa}$ - (a), $\rho_{bb}$ - (b), coherence $\text{Re}[\rho_{ab}]$ - (c), and $\text{Im}[\rho_{ab}]$- (d). Second row: 2D plots for the linear transmission signal $|S^{(1)}(\omega)|$ vs detected frequency $\omega$ and chirp rate $\phi''$ starting with initial population - (e), starting with coherence - (f) and the total signal - (g) for $\Omega=0.01$ eV and $\omega_0=0.035$ eV. Third row - the same as second row but for $\Omega=0.01$ eV and $\omega_0=0.05$ eV. Here $\omega_{ba}=0.05$ eV, $\omega_{ca}=1.0$ eV. Other parameters are the same as in Fig.~\ref{equilibrium}.}
\label{contour-phase2}
\end{figure}

In the following numerical calculations we first solve for the steady state $\tilde{\rho}_{ss}$ using the master equation given in Appendix B (see Eq.~(\ref{master-eq})) and then calculate the linear signal using Eq.~(\ref{eq:population}) and Eq.~(\ref{eq:coherence}). We use the linearly chirped Gaussian electric field ${\cal E}(\omega)$ given in Eq.~(\ref{chirp-pulse}). Fig \ref{equilibrium}(a) shows the
equilibrium case  when there is no driving i.e., $\omega_0=\Omega=0$. The signal then shows two peaks for
the dipole transitions between the states $|a\rangle \to |c\rangle$ and $|b\rangle \to
|c\rangle$. The peak at $\omega=\omega_{ac}$ is much stronger compared to the one at $\omega_{bc}$ since $\rho_{aa}-\rho_{cc} \gg\rho_{bb}-\rho_{cc}$ at equilibrium. Fig.~\ref{equilibrium}(b) depicts the static  situation with $\omega_0=0$ but finite $\Omega$. This static coupling between states $|a\rangle$ and
$|b\rangle$ renormalizes the energy values for these levels. These new energies are given by 
$\omega_a'=\frac{1}{2}(\omega_a+\omega_b) - \frac{1}{2} (4\Omega^2+ \omega_{ba}^2)^{\frac{1}{2}}$ and $\omega_b'=\frac{1}{2}(\omega_a+\omega_b)+\frac{1}{2} (4\Omega^2+ \omega_{ba}^2)^{\frac{1}{2}}$ with the gap between the two energy states is $(4\Omega^2 + \omega_{ba}^2)^{\frac{1}{2}}$ (see Fig. \ref{dressed} at $\omega_0=0$). Here $\omega_{ba}=\omega_b-\omega_a$. The signal shows two peaks at $\omega=\omega_c-\omega_b'$ and $\omega=\omega_c-\omega_a'$.

We next turn to the case of periodic driving with finite $\omega_0$. Fig.~\ref{bloch_signal2}(a) depicts the signal $S_{\text{pop}}^{(1)}(\omega)$ generated by the system prepared initially in the population state given by a field with  moderate strength $\Omega<\omega_0$. In the presence of a monochromatic field  the atomic wavefunction oscillates with four different frequencies \cite{Boyd} (see Fig. \ref{dressed}): $\omega_a-\frac{1}{2} (\Delta_{ab} \pm \Omega')$ and $\omega_b+\frac{1}{2} (\Delta_{ab} \pm \Omega')$ where  $\Delta_{ab}= \omega_0-\omega_{ba}$ and $\Delta_{ac}=\omega_0-\omega_{ca}$ are the detuning frequencies and $\Omega'=(4 \Omega^2 + \Delta_{ab}^2)^{\frac{1}{2}}$. To that end we observe four peaks in the signal which corresponds to following resonance frequencies  $\omega_{ca}-\frac{1}{2} (\Delta_{ab} \pm \Omega')$, and $\omega_{cb}+\frac{1}{2} (\Delta_{ab} \pm \Omega')$. Similarly we obtain the signal $S_{\text{coh}}^{(1)}(\omega)$ for the system initially prepared in coherence. Fig. \ref{bloch_signal2}(b) contains the same four peaks as in Fig. \ref{bloch_signal2}(a) but with the different intensity profile compared to the population contribution, due to different quantum pathways contributing to the signal. The overall intensity is suppressed, as coherence contribution is always weaker than the population. The total signal $S_{\text{tot}}^{(1)}(\omega)$ is depicted in Fig. \ref{bloch_signal2}(c) and is dominated by the population contribution. 

So far we discussed moderate driving $\Omega' < \omega_0$. In this case the gap between peaks 1 and 2 and between peaks 3 and 4 is $\Omega'$ whereas the gap between peak 1 and 3 and between 2 and 4 is $\omega_0$ (see Fig. \ref{dressed}). For strong driving $\Omega' > \omega_0$ due to the level crossing of the energy spectrum the gap between peak 1 and 2 and between 3 and 4 becomes $\omega_0$ whereas the gap between 1 and 3 and between 2 and 4 is $\Omega'$. The corresponding signals generated from the system initially prepared in the population state is shown in Fig. \ref{bloch_signal2}(d). The coherence contribution is shown in Fig. \ref{bloch_signal2}(e) and the total signal is plotted in Fig. \ref{bloch_signal2}(f). Strong driving results in a higher frequency resolution in the spectra where all four peaks are well separated in frequency. Furthermore,  the strong driving enhances the population contribution and suppresses the coherence contribution in the signal.
 
The effect of level crossing and dependence of the signal with respect to $\omega_0$ and $\Omega$ for a broad range of parameters is shown in Fig.~(\ref{contour}).  Fig.~\ref{contour}(a) depicts the absolute value of the signal $|S^{(1)}(\omega)|$ vs the detected frequency $\omega$ and the Rabi frequency $\Omega$ for fixed value of $\omega_0=0.02$ eV. The signal clearly shows four peaks with fixed splitting equal to $\omega_0$ between the nearest peaks (1 and 2) and (3 and 4). In the same time the gap between peak 1 and 3 and between 2 and 4 is $\Omega'$ which increases linearly with $\Omega$. Fig \ref{contour} (b) represents the signal vs $\omega$ and field frequency $\omega_0$ for fixed value of $\Omega=0.1$ eV. For small $\omega_0$ the separation between the the peaks are not well resolved. However for higher $\omega_0$ the separation between the nearest peaks increases linearly with $\omega_0$.

In Fig.~(\ref{contour-phase}) we display the chirp rate dependence of the linear signal. If the system initially prepared in the  population, the signal $S^{(1)}_{\rm pop}$ shows oscillatory behavior for all four peaks with respect to the chirp rate due to the error function in the field $\bar{\mathcal{E}}$.  The period of oscillations for the  (1 and 2) pair of states with high energy is slightly shorter than for the (3 and 4) pair of states with low energy. This is due to the since central frequency of the pulse ($\bar{\omega}_c=0.3$ eV) which results in the faster oscillation of $\omega_{ca}$ compared to $\omega_{cb}$.  This asymmetry is enhanced for the case when the system is prepared initially in coherence. Fig. \ref{contour-phase}(b) depicts the $S^{(1)}_{\rm coh}$, where in addition to error function, the nonlinear phase dependence comes from the oscillating exponent. However since, the total signal $S^{(1)}_{\rm tot}$ is dominated by the population contribution, the spectra in Fig.~\ref{contour-phase}(c) is nearly indistinguishable from the population contribution alone. Similar population dominating behavior  of the total signal was observed in Fig.~1. However unlike Fig.1 where we did not address the details of the preparation in a driven system the initial conditions can be controlled by the driving field. In fact the steady state populations and coherences depend differently on the parameters of the driving field \cite{Scully} such as $\Omega$ and $\omega_0$ which makes it possible, in principle, to control relative contributions of populations and coherence and extract a pure coherence contribution.

To demonstrate the effect of the control of the population/coherence contributions to the signal we examine first the steady state density matrix elements as a function of $\Omega$ and $\omega_0$. Fig. \ref{contour-phase2}(a) and (b) show the steady-state populations $\rho_{aa}$ and $\rho_{bb}$, respectively as a function of $\omega_0$ for the fixed value of the Rabi frequency $\Omega$. For a given parameters $\rho_{aa}$ ($\rho_{bb}$) are symmetric functions of its argument which reaches its minimum (maximum) at the resonance frequency $\omega_0=0.05$ eV. Both populations are concentrated around the value of $0.5$ as the total populations of all states is always unity and $\rho_{cc}=0$ in the steady state. The same applies to the imaginary part of the coherence $\text{Im}[\rho_{ab}]$ as shown in Fig. \ref{contour-phase2}(d) with the value changing from $-0.01$ to $-0.07$. On the other hand the real part of the coherence $\text{Re}[\rho_{ab}]$ is an asymmetric function of its argument ranging from $0.2$ to $-0.2$ and reaches zero at $\omega_0=0.05$ with the maximum value around $\omega_0=0.035$ eV and $\omega_0=0.055$ eV. Therefore we have identified two parameter regimes when coherence contribution is substantial with the steady state value of $0.2$ ($\Omega=0.01$ eV, $\omega_0=0.035$ eV - parameter regime I) and regime when it can be neglected ($\Omega=0.01$ eV, $\omega_0=0.05$ eV - parameter regime II). Fig. \ref{contour-phase2}(e) shows the oscillations of the population contribution with respect to nonlinear phase $\phi''$ using  for parameter regime I. Due to weak driving $\Omega=0.01$ not all four peaks depicted previously in Fig. \ref{contour-phase}  are well resolved. In fact there is one strong peak at $\omega=0.98$ eV and two weak peaks at $\omega=0.94$ eV and $\omega=1.0$ eV. Coherence contribution to the signal is shown in Fig. \ref{contour-phase2}(f). In this case all three peaks are manifested much stronger. Finally for the total signal shown in Fig. \ref{contour-phase2}(g) we see that out of three peaks the strong peak at $\omega=0.98$ eV and  weak peak at $\omega=0.96$ are suppressed whereas the weak peak at $\omega=1.0$ eV is enhanced compared to the pure population contribution shown in Fig. \ref{contour-phase2}(e). This is a manifestation of the strong coherence. In the parameter regime II the field frequency is $\omega=0.05$ eV and the corresponding value of the steady state coherence drops by roughly a factor of 3. In this case the population contribution shown in Fig. \ref{contour-phase2}(h) dominates over the coherence contribution shown Fig. \ref{contour-phase2}(i) and the total signal in Fig. \ref{contour-phase2}(j) is completely dominated by the population contribution. Therefore, a driven preparation allows to selectively manipulate the initial conditions of the system and separate the quantum pathways contributions from initial populations and coherences to the linear absorption signal. Note that similar analysis can be performed if we instead fix the value of $\omega_0$ and change the intensity of the driving field via $\Omega$.



\section{Conclusions} 
We have calculated the  
linear and non-linear frequency domain optical signals for systems prepared in a nonequilibrium state. The generalized $n$-th order susceptibility $\tilde{\chi}^{(n)}$ then depends on
$n\!+\!1$ independent frequency variables rather than $n$, which is the case for systems initially in equilibrium. This nonequilibrium state results in a nontrivial phase dependence of the electric field, already in the linear signal, even if the system is initially in a population state. This phase dependence is strongly enhanced by initial coherences. Furthermore we predict the new resonances in nonlinear signals that depend on the frequency difference between the initially prepared coherent superposition of molecular states. We then addressed a particular case of the initial preparation of the system. We investigated a three level system with two lower energy states strongly driven by a monochromatic field. In this case the driving field frequency $\omega_0$ generates a phase dependent linear signal. Performing numerical simulations we show that depending on the strength of the driving field via its Rabi frequency, the relative contributions to the signals due to initial populations and coherences can be controlled and even a pure coherence contribution can be extracted.

The new resonances and phase dependence demonstrated in the present work are closely related to quantum coherences.  For instance, the atomic gas strongly driven with microwave radiation allows to alter detailed balance conditions and change the transmission properties of the optical pulses under certain conditions for electronic transitions and parameters of the driving and probe fields. These effects have been studied in cold alkali gases where effects such as lasing without population inversion \cite{Harris,Lasing1, Lasing2}, electromagnetically induced transparency \cite{Fle05}, and slow light \cite{Hau99,Kas99} were observed. These effects have also been observed in semiconductor quantum dots \cite{Cho03,Cho07}, and heterostructures  \cite{Bel01,Dor12}. These fundamental effects are typically examined through simple linear transmission/absorption experiments well described by our formalism. Furthermore, the parameter regime for observation of coherence effects is very similar to the one used in the simulations in Figs.~5-10. The power of the approach presented in this paper is not limited to linear experiments. The results of Eqs.~(26)-(31) are described in terms of Liouville space electronic Green's functions and thus allow to extend quantum coherence effects to higher order nonlinear optical measurements exploring interplay between coherence and many-body effects in atomic, molecular and solid state systems. These include electron and nuclear dynamics, nonadiabatic dynamics, chemical reactions and other many-body effects. We examined a series of measurements and their understanding in the context of spectroscopy of e.g. strongly driven systems. We demonstrated how electronic and vibrational structure and dynamics of molecules influenced by strong driving fields. The practical applications of our formalism in the context of quantum coherence effects discussed above has great potential in solid state technology. In particular, in the recent years quantum coherence in nitrogen vacancy (NV) defect centers in diamond show great promise for quantum information processing \cite{Jel06}, magnetometry \cite{Maz08,Mal12,Ron12} and electrometry \cite{Dol11}. The recently measured temperature dependence of the zero-field splitting constant \cite{Aco10,Toy12} indicates that it may also be used as an atomic temperature sensors. The broadband excitation of the NV centers in diamond with chirped pulses \cite{Nie13} allow the quantum control of the phases of single electron spins. The phase dependence of the optical signals shown by authors in Fig.~5 of Ref.~\cite{Nie13} yields similar features to our Fig.~7 of the present paper. The predicted resonances and phase control along with nonlinear optical measurements can be tested in these solid state systems.

\section{Acknowledgement}
We want to thank Kochise Bennett and Arunangshu Debnath for useful and stimulating
discussions. We gratefully acknowledge the support of National Science Foundation (NSF)
through grant no.~CHE-1361516, US Department of Energy (DOE) and the National Institute of
Health (NIH) grant no. GM-59230.
 
\appendix

\section*{Appendix A: Four wave mixing signals} 

\begin{figure}
\includegraphics[width=0.7\columnwidth]{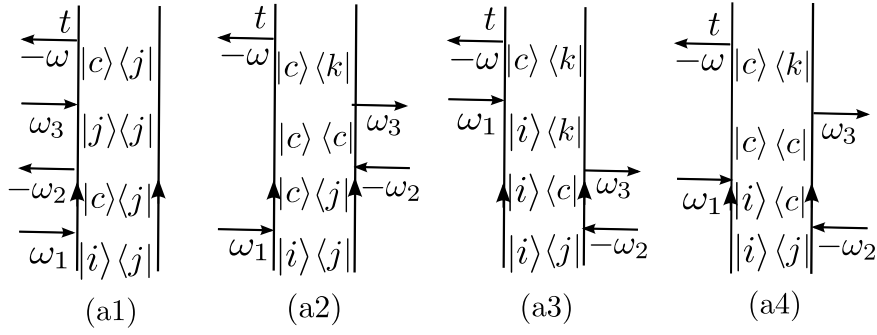}
\caption{Ladder diagrams for three level system shown in Fig.~(\ref{level-scheme-plots}a). Here the states $|i\rangle, |j\rangle,|k\rangle$ representes two lower energy states $|a\rangle$ and $|b\rangle$ of three level system.} 
\label{pathways-FWM}
\end{figure}

In Fig.~(\ref{pathways-FWM}) we draw all possible Liouville space pathways (for both initial populations and initial coherences) that contribute to FWM signal for the three level model system in Fig.~(\ref{level-scheme-plots})a. For diagrammatic rules we refer to Ref.~\cite{Harbola}. The signal is calculated using Eq.~(\ref{FWM-eq}) considering three incoming fields ${\cal E}_i, i=1,2,3$ as monochromatic continuous waves and the probe field ${\cal E}_4$ as a spectrally broad pulses.  
The expression for the signal for these pathways is given as 
\bea
S_{\text{a1}}^{(3)}(\omega;\omega_1,\omega_2,\omega_3)&\!=\!& \frac{2}{\hbar} \,{\rm Re} \sum_{i,j,k \in a,b} \rho_{ij} \Bigg[  
\frac{{\cal E}_{4}^{*}(\omega) {\cal E}_3 {\cal E}_2^{*} {\cal E}_1}{(\omega\!-\!\omega_1\!+\!\omega_2\!-\!\omega_3\!-\!\omega_{ij}+i \eta)} \times \frac{\mu_{jc} \mu_{kc}^* \mu_{kc} \mu_{ic}^*}{(\omega\!-\!\omega_{cj}\!+\!i \eta)(\omega\!-\!\omega_3\!-\!\omega_{kj}\!+\!i \eta)} \nonumber \\
&&\times \frac{1}{\omega\!+\!\omega_2\!-\!\omega_3\!-\!\omega_{cj}\!+\!i\eta}\Bigg] + (\omega_1 \leftrightarrow \omega_3)
\label{path1}
\eea
\bea
S_{\text{a2}}^{(3)}(\omega;\omega_1,\omega_2,\omega_3)&\!=\!& \frac{2}{\hbar} \,{\rm Re} \sum_{i,j,k \in a,b} \rho_{ij} \Bigg[  
\frac{{\cal E}_{4}^{*}(\omega) {\cal E}_3 {\cal E}_2^{*} {\cal E}_1}{(\omega\!-\!\omega_1\!+\!\omega_2\!-\!\omega_3\!-\!\omega_{ij}+i \eta)} \times \frac{\mu_{kc} \mu_{cj} \mu_{kc}^* \mu_{ic}^*}{(\omega\!-\!\omega_{ck}\!+\!i \eta)(\omega\!-\!\omega_3\!\!+\!i \eta)} \nonumber \\
&&\times \frac{1}{\omega\!+\!\omega_2\!-\!\omega_3\!-\!\omega_{cj}\!+\!i\eta}\Bigg] + (\omega_1 \leftrightarrow \omega_3)
\label{path2}
\eea
\bea
S_{\text{a3}}^{(3)}(\omega;\omega_1,\omega_2,\omega_3)&\!=\!& \frac{2}{\hbar} \,{\rm Re} \sum_{i,j,k \in a,b} \rho_{ij} \Bigg[  
\frac{{\cal E}_{4}^{*}(\omega) {\cal E}_3 {\cal E}_2^{*} {\cal E}_1}{(\omega\!-\!\omega_1\!+\!\omega_2\!-\!\omega_3\!-\!\omega_{ij}+i \eta)} \times \frac{\mu_{kc} \mu_{cj} \mu_{kc}^* \mu_{ic}^*}{(\omega\!-\!\omega_{ck}\!+\!i \eta)(\omega\!-\!\omega_1\!-\!\omega_{ik}\!+\!i \eta)} \nonumber \\
&&\times \frac{1}{\omega\!-\!\omega_1\!-\!\omega_3\!-\!\omega_{ic}\!+\!i\eta}\Bigg] + (\omega_1 \leftrightarrow \omega_3)
\label{path3}
\eea
\bea
S_{\text{a4}}^{(3)}(\omega;\omega_1,\omega_2,\omega_3)&\!=\!& \frac{2}{\hbar} \,{\rm Re} \sum_{i,j,k \in a,b} \rho_{ij} \Bigg[  
\frac{{\cal E}_{4}^{*}(\omega) {\cal E}_3 {\cal E}_2^{*} {\cal E}_1}{(\omega\!-\!\omega_1\!+\!\omega_2\!-\!\omega_3\!-\!\omega_{ij}+i \eta)} \times \frac{\mu_{kc} \mu_{cj} \mu_{kc}^* \mu_{ic}^*}{(\omega\!-\!\omega_{cj}\!+\!i \eta)(\omega\!-\!\omega_3\!\!+\!i \eta)} \nonumber \\
&&\times \frac{1}{\omega\!-\!\omega_1\!-\!\omega_3\!-\!\omega_{ic}\!+\!i\eta}\Bigg] + (\omega_1 \leftrightarrow \omega_3)
\label{path4}
\eea
The total signal is $S_{\text{total}}^{(3)}=S_{\text{a1}}^{(3)}+S_{\text{a2}}^{(3)}+S_{\text{a3}}^{(3)}+S_{\text{a4}}^{(3)}$.

\section*{Appendix B : Master equation for a driven three level system and the linear response signal}
In this section we give the details about the master equation for the driven three level system described by the Hamiltonian given in Eq.~(\ref{Driven-Bloch-1}). We follow the standard set of approximations such as weak system-bath coupling (second
order), large reservoirs, wide-band of the leads \cite{Scully, Agarwal, Boyd, Harbola} to derive the 
equations of motions for
populations and coherences. 

We Define new set of variables (denoted by the symbol tilde) in the rotating frame $\tilde{\rho}_{ab}(t)= e^{-i\omega_0
t}{\rho}_{ab}(t)$, $\tilde{\rho}_{ac}(t)= e^{-i \omega_0
t}{\rho}_{ac}(t)$,
$\tilde{\rho}_{bc}=\rho_{bc}$ and 
$\tilde{\rho}_{ii}=\rho_{ii}, i=a,b,c$, to obtain the Bloch equation
\bea
\tilde{\dot{\rho}}_{aa}&=&i \Omega (\tilde{\rho}_{ba}-\tilde{\rho}_{ab}) -
(\gamma_{ab}+\gamma_{ac})\tilde{\rho}_{aa} + \gamma_{ba} \tilde{\rho}_{bb} + \gamma_{ca}
\tilde{\rho}_{cc} \nonumber \\
\tilde{\dot{\rho}}_{bb}&=&-i \Omega (\tilde{\rho}_{ba}-\tilde{\rho}_{ab}) 
-(\gamma_{ba}+\gamma_{bc})\tilde{\rho}_{bb} + \gamma_{ab} \tilde{\rho}_{aa} + \gamma_{cb}
\tilde{\rho}_{cc} \nonumber \\
\tilde{\dot{\rho}}_{ab}&=& -i \Delta_{ab} \tilde{\rho}_{ab} +i \Omega (\tilde{\rho}_{bb}- \tilde{\rho}_{aa})
-\frac{1}{2}(\gamma_{ab}+\gamma_{ba}+\gamma_{ac}+\gamma_{bc})\tilde{\rho}_{ab} \nonumber
\\
\tilde{\dot{\rho}}_{ac}&=& -i \Delta_{ac} \tilde{\rho}_{ac} +i \Omega \tilde{\rho}_{bc} 
-\frac{1}{2}(\gamma_{ac}+\gamma_{ca}+\gamma_{ab}+\gamma_{cb})\tilde{\rho}_{ac} \nonumber
\\
\tilde{\dot{\rho}}_{bc}&=& -i (\omega_b-\omega_c)\tilde{\rho}_{bc} +i  \Omega
\tilde{\rho}_{ac}
-\frac{1}{2}(\gamma_{bc}+\gamma_{cb}+\gamma_{ba}+\gamma_{ca})\tilde{\rho}_{bc}
\label{master-eq}
\eea
where $\gamma_{ij}, i,j=a,b,c$
are the decay rates from the state $i$ to state $j$. Note that, the driving field couples the population of the lower two states with its coherence. The crucial advantage of working with the rotating frame is that the Liouville operator in this frame $\tilde{\cal L}$ becomes time-independent and the steady-state solution can be obtained. We write the above equation in matrix form as $\dot{\tilde{\rho
}}(t)= \tilde{\cal L} \tilde{\rho}(t)$ where
$\tilde{\rho}^T= (\tilde{\rho}_{aa},\tilde{\rho}_{bb}, \tilde{\rho}_{cc},
\tilde{\rho}_{ab}, \tilde{\rho}_{ba}, \tilde{\rho}_{ac}, \tilde{\rho}_{ca},
\tilde{\rho}_{bc}, \tilde{\rho}_{cb})$. The transformation matrix between lab and
rotating frame is diagonal
\be
{U}(t) = {\rm diag} (1,1, 1, e^{i \omega_0 t}, e^{-i \omega_0 t}, e^{i \omega_0 t},
e^{-i \omega_0 t}, 1 , 1)
\ee
where diag represents the diagonal elements of the matrix. Since $\rho(t) = {U}(t) \tilde{\rho}(t)$ it implies that the propagators in the two frames are related via
\be
{\cal G}(t,t')= { U}(t) \,\tilde{{\cal G}}(t-t')\, { U}^{-1}(t')
\label{G-relation}
\ee
where $\tilde{\cal G}(t,t')= -\frac{i}{\hbar} \theta(t-t') e^{\tilde{\cal L}(t-t')}$ and ${\cal
G}(t,t')= -\frac{i}{\hbar}  \theta(t-t') \,T e^{\int_{t'}^{t}d\tau {\cal L}(\tau) }$ for $t>t'$.
We first solve for the steady state in the rotating frame by demanding $\dot{\tilde{\rho}}_{ss}=0$.
Then in the lab frame the solution for non-stationary state $\rho_{nss}$ at a particular time $\tau_0$ (taken as the preparation time) is given as $\rho_{nss}(\tau_0)=
U(\tau_0) \tilde{\rho}_{ss}$ which also mean that the coherence elements oscillate with the driving field frequency $\omega_0$ even in the long-time limit.

We probe the system, prepared in the state $\rho_{nss}(\tau_0)$, with a weak probe field
${\cal E}(t)$ which allows transitions between the states $|a\rangle \to |c\rangle$ and $|b\rangle \to |c\rangle$. We impose the RWA and write the light-matter interaction Hamiltonian as
\be
H_{\rm int}(t)= {\cal E}(t) V^{\dagger}+{\cal E}^*(t) V
\ee
where $V= \sum_{i=a,b} \mu_{ic} |i\rangle \langle c|$. Given the initial state
$\rho_{nss}(\tau_0)$ we write the expression for the linear
signal following Eq.~(\ref{expand-first}) as 
\be
S^{(1)}(\omega;\omega_0,\Omega)= \frac{2}{\hbar}\,{\rm Im}\Bigg[i \hbar \, {\cal E}^{*}(\omega)
\int_{-\infty}^{\infty} dt
e^{i\omega
t} \int_{-\infty}^{t} d\tau_1 \,{\cal E}(\tau_1)\, {\rm Tr} \big[V_L {\cal G}(t,\tau_1)
V_{-}^{\dagger} {\cal G}(\tau_1-\tau_0) \rho_{nss}(\tau_0)\big]\Bigg] 
\ee
Using the relation between the propagators ${\cal G}(t,t')$ and $\tilde{{\cal G}}(t-t')$ given in Eq.~(\ref{G-relation}) we obtain
\be
S^{(1)}(\omega;\omega_0,\Omega)=\frac{2}{\hbar}\,{\rm Im}\Bigg[{\cal E}^{*}(\omega)
\int_{-\infty}^{\infty} dt
e^{i\omega
t} \int_{-\infty}^{t} d\tau_1 \,{\cal E}(\tau_1) \langle V_L { U}(t)
\tilde{\cal G}(t-\tau_1) { U}^{-1}(\tau_1) V_{-}^{\dagger} {
U}(\tau_1)\rangle_{\tilde{\rho}_{ss}} \Bigg] \quad \quad 
\ee
Note that in the last line the average is expressed with respect to $\tilde{\rho}_{ss}$ which is obtained by using the relation $\tilde{\cal G}(\tau_1-\tau_0) \tilde{\rho}_{ss}=-\frac{i}{\hbar} \tilde{\rho}_{ss}$. Performing Fourier transformation for the field and the propagator and writing down the matter correlation function explicitly by reading the diagrams for population and coherence in Fig.~\ref{ladder-diagrams} we obtain Eq.~(\ref{eq:population}) and Eq.~(\ref{eq:coherence}).


\begin{thebibliography}{99}

\bibitem{Shaul5} S. Mukamel, '{\it Principles of Nonlinear Optical Spectroscopy}' (Oxford
Univ. Press, New York).

\bibitem{FWM_Mukamel} S. Mukamel, R. F. Loring, J. Opt. Soc. Am. B {\bf 3}, 595-606
(1986).

\bibitem{Shaul2} S. Mukamel, Phys. Rev. E {\bf 68}, 021111 (2003). 


\bibitem{Miz97} Y, Mizutani and T. Kitagawa, Science {\bf 278}, 443 (1997).
\bibitem{Lee:JCP:2004} S.-Y. Lee, D. Zhang, D. W. McCamant, P. Kukura, and R. A. Mathies, J. Chem. Phys {\bf 121}, 3632 (2004).
\bibitem{Kukura:AnnurevPhysChem:2007} P. Kukura, D. W. McCamant and R. A. Mathies, Annu. Rev. Phys. Chem., {\bf 58}, 461 (2007).
\bibitem{Takeuchi:Science:2008} S. Takeuchi, S. Ruhman, T. Tsuneda, M. Chiba, T. Taketsugu, and T. Tahara, Science {\bf 322}, 1073 (2008).

\bibitem{Scully} M. O. Scully, M. S. Zubairy, '{\it Quantum Optics}' (Cambridge Univ Press, Cambridge,
England) (1997).
\bibitem{Boyd} R. W. Boyd, '{\it Nonlinear optics}'. Academic Press, (2003).
\bibitem{Boukobza} E. Boukobza and D. J. Tannor, Phys. Rev. Lett, {\bf 98}, 240601 (2007).

\bibitem{Geva} E. Geva and R. Kosloff, Phys. Rev. E, {\bf 49}, 3903 (1994). 
\bibitem{Scovil} H. E. D. Scovil and E. O. Schulz-DuBois, Phys. Rev. Lett, {\bf 2}, 262 (1959).
\bibitem {Harris} S. E. Harris,  Phys Today {\bf 50} 36. (1997).
\bibitem{Lasing1} O. Kocharovskaya Phys Rep {\bf 219} 175 (1992).
\bibitem{Lasing2} S. Ya. Kilin, K. T. Kapale and M. O. Scully, Phys. Rev. Lett. 100, 173601 (2008).
\bibitem{CPT1} G. Alzetta, A. Gozzini, L. Moi and G. Orriols, Nuovo Cimento, 36B, 5 (1976).
\bibitem{CPT2} E. Arimondo and G. Orriols, Nuovo Cimento Lett. 17, 333 (1976).
\bibitem{FD1} U.M.B. Marconi, A. Puglisi, L. Rondoni, A. Vulpiani,  Phys. Rep. {\bf 461}, 111 (2008).
\bibitem{FD2} T. Harada and S. I. Sasa, Phys. Rev. Lett {\bf 95}, 130602 (2005).
\bibitem{FD3} K. Saito, Europhys. Lett. {\bf 83}, 50006 (2008).





\bibitem{control1} V.I. Prokhorenko, A. M. Nagy, S. A. Waschuk, L. S. Brown, R. R. Birge,
and R. J. D. Miller, Science {\bf 313}, 1257 (2006).

\bibitem{control2} M. Spanner, C. A. Arango, and P. Brumer, J. Chem. Phys. {\bf 133}, 151101 (2010).

\bibitem{Shaul1} S. Mukamel, J. Chem. Phys. {\bf 139}, 164113 (2013).
\bibitem{Rahav_stim} S. Rahav, O. Roslyak, and S. Mukamel, J. Chem. Phys. {\bf 131}, 194510
(2009).
\bibitem{Rahav2_stim} S. Rahav and S. Mukamel, J. Chem. Phys. {\bf 133}, 244106 (2010).
\bibitem{Pan10} G. Panitchayangkoon, D. Hayes, K.A. Fransted, J.R. Caram, E. Harel, J. Wen, R.E. Blankenship, G.S. Engel, Proc. Natl. Acad. Sci. \textbf{107}, 12766 (2010).
\bibitem{Pro14} F. Provencher, N. Berube, A.W. Parker, G.M. Greetham. M. Towrie, C. Hellman, M. Cote, N. Stingelin, C. Silva and S.C. Hayes, Nat. Comm. \textbf{5}, 4288 (2014).



\bibitem{chirp} B. Chatel,J. Degert, and B. Girard, Phys. Rev. A {\bf 70}, 053414 (2004).
\bibitem{Debnath} A. Debnath, C. Meier, B. Chatel, and T. Amand, Phys. Rev. B {\bf 86}, 161304(R) (2012). 











\bibitem{Harbola} C. Marx, U. Harbola, S. Mukamel, Phys. Rev. A {\bf 77}, 022110 (2008).




\bibitem{Fle05}M. Fleischhauer, A. Imamoglu, and J. P. Marangos, Rev. Mod. Phys. {\bf 77}, 633 (2005).
\bibitem{Hau99}L. V. Hau, S. E. Harris, Z. Dutton, and C. H. Behroozi, Nature {\bf 397}, 549 (1999).
\bibitem{Kas99} M. M. Kash,V. A. Sautenkov, A. S. Zibrov, L.Hollberg, G. R. Welch, M. D. Lukin,Y.Rostovtsev, E. S. Fry, and M. O. Scully, Phys. Rev. Lett. {\bf 82}, 5229 (1999).
\bibitem{Cho03} W. W. Chow, H. C. Schneider, and M. C. Phillips, Phys. Rev. A {\bf 68}, 053802 (2003). \bibitem{Cho07} W. W. Chow, S. Michael, and H. C. Schneider, J. Mod. Opt. {\bf 54}, 2413 (2007).
\bibitem{Bel01} A. A. Belyanin, F. Capasso, V. V. Kocharovsky, Vl. V. Kocharovsky, and M. O. Scully, Phys. Rev. A {\bf 63}, 053803 (2001).
\bibitem{Dor12} K.E. Dorfman, M.B. Kim, and A.A. Svidzinsky, Phys. Rev. A {\bf 84}, 053829 (2011).

\bibitem{Jel06} F. Jelezko and J. Wrachtrup, Phys. Status Solidi a {\bf 203}, 3207Ð25 (2006).
\bibitem{Maz08} J.R. Maze et al Nature {\bf 455} 644Ð7 (2008).
\bibitem{Mal12} P. Maletinsky, S. Hong, M.S. Grinolds, B. Hausmann,  M.D. Lukin, R.L. Walsworth, M. Loncar and A. Yacoby, Nature Nano {\bf 7} 320Ð4 (2012).
\bibitem{Ron12} L. Rondin, J.-P. Tetienne, P. Spinicelli, C. Dal Savio, K. Karrai, G. Dantelle, A. Thiaville, S. Rohart, J.-F. Roch J-F and V. Jacques, Appl. Phys. Lett. {\bf 100} 153118 (2012).
\bibitem{Dol11} F. Dolde et al, Nature Phys. {\bf 7} 459Ð63 (2011).
\bibitem{Aco10} V.M. Acosta, E. Bauch, M.P. Ledbetter, A. Waxman, L.-S. Bouchard and D. Budker, Phys. Rev. Lett. {\bf 104} 070801 (2010).
\bibitem{Toy12} D.M. Toyli, D.J. Christle, A. Alkauskas, B.B. Buckley, C.G. Van de Walle and D.D. Awschalom, Phys. Rev. X {\bf 2} 031001 (2012).
\bibitem{Nie13} I. Niemeyer, J.H. Shim, J. Zhang, D. Suter, T. Taniguchi, T. Teraji, H. Abe, S. Onoda, T. Yamomoto, T. Ohshima, J. Isoya, and F. Jelezko, New J. Phys. {\bf 15}, 033027 (2013).
\bibitem{Agarwal} G. Agarwal, '{\it  Quantum Statistical Theories of Spontaneous Emission and
Their Relation to Other Approaches}', Springer Tracts in Modern Physics (Springer, Berlin),
Vol 70 (1974).

\end{thebibliography}

\end{document}